\documentclass[a4paper,prd,preprintnumbers,,twocolumn,superscriptaddress,nofootinbib,amsmath,amssymb]{revtex4}
\usepackage[dvips]{graphics}
\usepackage{color,hyperref}
\usepackage{times}
\usepackage{mathrsfs}
\hypersetup{colorlinks=true,linkcolor=blue,citecolor=red,filecolor=magenta,urlcolor=blue}
\usepackage{enumitem}
\usepackage{graphicx}
\usepackage{subfig}
\usepackage[utf8]{inputenc}
\usepackage{dcolumn}
\usepackage{subcaption} 
\usepackage{bm}
\usepackage{color}
\usepackage{xcolor}

\begin{document}

\title{Gravitational lensing effects on neutrino oscillations around a Schwarzschild black hole immersed in perfect fluid dark matter}


\author{Sojida Mannobova}
\email{sojidamannabova2208@gmail.com}

\affiliation{National University of Uzbekistan, Tashkent 100174, Uzbekistan}

\author{Farruh Atamurotov}
\email{atamurotov@yahoo.com}
\affiliation{Kimyo International University in Tashkent, Shota Rustaveli str. 156, Tashkent 100121, Uzbekistan}
\affiliation{Research Center of Astrophysics and Cosmology, Khazar University, 41 Mehseti Street, Baku AZ1096, Azerbaijan}

\author{Ahmadjon~Abdujabbarov}
\email{ahmadjon@astrin.uz}

\affiliation{School of Physics, Harbin Institute of Technology, Harbin 150001, People’s Republic of China}
\affiliation{Institute of Fundamental and Applied Research, National Research University TIIAME, Kori Niyoziy 39, 100000 Tashkent, Uzbekistan}

\author{Phongpichit Channuie}
\email{phongpichit.ch@mail.wu.ac.th}
\affiliation{School of Science, Walailak University, Nakhon Si Thammarat, 80160, Thailand}
\affiliation{College of Graduate Studies, Walailak University, Nakhon Si Thammarat, 80160, Thailand}

\author{Chengxun Yuan}
\email{yuancx@hit.edu.cn}
\affiliation{School of Physics, Harbin Institute of Technology, Harbin 150001, People’s Republic of China}

\author{Chen Zhou
    }
\email{chenzhou@hit.edu.cn}
\affiliation{School of Physics, Harbin Institute of Technology, Harbin 150001, People’s Republic of China}

\begin{abstract}
We investigate neutrino flavor oscillations in the weak gravitational-lensing regime of a static, spherically symmetric black hole surrounded by perfect fluid dark matter (PFDM). Within a covariant quantum-mechanical framework, we derive the oscillation phases for radial and non-radial propagation of ultrarelativistic neutrinos. While the leading radial phase is independent of the metric function, non-radial trajectories acquire impact-parameter-dependent corrections governed by the PFDM parameter $\lambda$. We construct the flavor-transition probability for two coherent lensed paths and analyze the resulting interference in two and three flavor scenarios. Varying $\lambda$ shifts the oscillation extrema and modifies their amplitudes and contrast, with a non-monotonic dependence on the neutrino mass scale, lens mass, flavor channel, and mass ordering. The $P_{\mu\tau}$ channel shows particularly strong sensitivity to the PFDM contribution for the considered configurations. Moreover, the path-interference terms retain information about the absolute neutrino mass scale. These results characterize PFDM-induced modifications of lensed neutrino oscillations and provide a basis for more realistic treatments including decoherence, averaging, detector resolution, and corrections to lens trajectories and magnifications.

\end{abstract}

\maketitle

\section{Introduction}

Astrophysical black holes are not isolated vacuum systems. They reside in galaxies, accrete baryonic matter, and may be embedded in dark-matter distributions~\cite{Gondolo:1999ef,Cardoso:2021wlq} whose gravitational field contributes to the geometry sampled by particles and radiation. Although the cosmological and galactic evidence for dark matter is
compelling~\cite{Bertone:2004pz,Planck:2018vyg}, its microscopic identity and its distribution near compact objects remain uncertain. This uncertainty motivates phenomenological geometries in which an ambient dark component is incorporated into the gravitational source.

One useful example is the perfect-fluid-dark-matter (PFDM) construction, originally introduced to account for the flat rotation curves of spiral galaxies~\cite{Rahaman:2010xm,LiYang2012}, for which the static metric acquires a logarithmic correction controlled by a parameter characterizing the surrounding matter distribution. Related solutions have been developed for rotating black holes and for black holes embedded in more general dark-matter halos~\cite{Xu:2017bpz,Xu:2018mkl}. In PFDM backgrounds specifically, the surrounding matter can alter particle motion, and quasiperiodic oscillations~\cite{Shaymatov:2021pfdm,Das:2021otl,Narzilloev22c,Mustafa:2024cfe}, the size of the black-hole shadow, which has been confronted with Event Horizon Telescope observations~\cite{Hou:2018bar,Haroon:2018ryd,Atamurotov:2021hck,Atamurotov:2024nre}, and the deflection of light, including in the presence of plasma~\cite{Atamurotov:2021hoq,Shodikulov:2025xax}. Such models provide a tractable means of asking whether environmental gravity can leave distinguishable signatures in observables usually employed to test the compact object itself.

This question has become increasingly relevant because black-hole observations now cover several independent dynamical regimes. Gravitational-wave measurements constrain the motion and merger of compact binaries, whereas horizon-scale imaging probes null geodesics and the radiative environment near supermassive black holes~\cite{Ligo2016,Akiyama19L1,2022ApJ...930L..12E}. At the
theoretical level, timelike orbits determine circular motion, stability boundaries, and characteristic orbital frequencies, while null trajectories govern light bending, photon rings, and the apparent shadow~\cite{Chandrasekhar:1985kt,Perlick:2021aok}. Consequently, particle dynamics, accretion observables, quasiperiodic oscillations, shadows, and weak or strong lensing constitute complementary tests of
both nonvacuum environments and alternative compact-object
metrics~\cite{Bambi:2015kza,Cardoso:2019rvt,Psaltis:2018xkc,Bozza:2002zj,2025arXiv250616405M}. Because environmental corrections can imitate or obscure changes attributed to non-Schwarzschild geometry~\cite{Barausse:2014tra}, observables based on physically different messengers are needed to
disentangle these effects.

Neutrinos offer a particularly different channel. They are produced and detected as flavor states, whereas propagation is described by mass eigenstates related through the Pontecorvo-Maki-Nakagawa-Sakata mixing matrix~\cite{Pontecorvo:1957qd,Maki:1962mu,Pontecorvo:1967fh}. The relative quantum phases accumulated by these eigenstates generate flavor oscillations, which in ordinary vacuum measurements determine mixing angles and mass-squared
differences~\cite{Bilenky:1987ty,Esteban:2024eli} but not the common absolute mass scale, which is instead bounded by direct kinematic measurements~\cite{KATRIN:2024cdt}. Since neutrinos interact only weakly with matter, they can escape from dense production regions and preserve
information complementary to electromagnetic
signals~\cite{Kamiokande-II:1987idp,Bionta:1987qt}. Their phases nevertheless respond to spacetime curvature. The gravitationally induced oscillation phase was first considered by Stodolsky~\cite{Stodolsky:1978ks}, revisited for weak static fields~\cite{Ahluwalia1996}, and subsequently analyzed in the Schwarzschild
geometry~\cite{Piriz:1996mu,Cardall:1996cd,Fornengo:1996ef,PhysRevD.59.067301,Pereira:2000kq}.
These studies were later extended to rotating
spacetimes~\cite{Lambiase_2005,Ren:2010yf,Visinelli:2014xsa} and to deformed or modified-gravity
backgrounds~\cite{Boshkayev2020,Buoninfante:2019der}, using both quantum-mechanical and field-theoretic descriptions.

Gravitational lensing enriches this phase evolution by introducing more than one classical route between the emission and detection events. If the wave packets associated with the relevant trajectories overlap
coherently at the detector, their amplitudes must be added before the flavor probability is evaluated~\cite{Swami:2021wkp}. The detected
pattern then combines the interference among the mass eigenstates with the interference among the propagation
paths~\cite{Crocker:2003cw,Alexandre:2018crg}. Unlike the conventional single-path vacuum probability, the lensed probability can contain combinations of the individual mass squares and may consequently depend on the absolute mass scale and, in suitable configurations, on the mass
ordering~\cite{Swami:2020qdi}. This mechanism has been explored for deformed and parametrized compact
objects~\cite{Chakrabarty:2021bpr,2024ChPhC..48g3103A,Alloqulov:2025param}, for black holes in modified and higher-dimensional gravity~\cite{Wang:2025nyl,Chakrabarty:2023kk}, for quantum-corrected geometries~\cite{AraujoFilho:2024mvz,Abdukayumova:2026pkn,Mannobova:2026jna}, and for wormhole spacetimes~\cite{Shi:2024wormhole}. It also makes
neutrino interference potentially sensitive to matter surrounding a lens, because an environmental contribution changes the null trajectories and their relative phases even when its direct effect on radial propagation is absent or
suppressed~\cite{Alloqulov:2025edn,Ergashov:2026mss}.

In the present work, we study flavor oscillations of neutrinos propagating in the spacetime of a static, spherically symmetric black hole immersed in PFDM. We derive the covariant phase for radial and nonradial motion and confirm that, at the order considered, the PFDM
term cancels from the radial phase but enters the
impact-parameter-dependent phase relevant to lensing. Going beyond previous two-flavor treatments, we construct the coherent two-path transition probability in both two- and three-flavor scenarios and investigate how the PFDM parameter $\lambda$ changes its angular
interference structure, including the dependence on the neutrino mass scale, mass ordering, flavor channel, and lens mass. 

The article is organized as follows. Section~\ref{sec2} introduces the static black-hole geometry surrounded by PFDM and specifies the weak-field domain used in the calculation. Section~\ref{sec3} reviews the covariant oscillation formalism. In Sec.~\ref{sec4} we obtain the phases for radial and nonradial propagation, and Sec.~\ref{sec5} develops the general coherent lensing probability. Section~\ref{sec6} presents the weak-lensing configuration and the numerical results for two and three neutrino flavors. The main conclusions and the limitations of the coherent approximation are summarized in Sec.~\ref{sec7}. Unless otherwise stated, we employ geometrized units $G=c=1$ and the metric signature $(-,+,+,+)$.

\section{BLACK HOLE IMMERSED IN PERFECT FLUID DARK MATTER \label{sec2}}

We consider a static and spherically symmetric black hole immersed in a 
perfect fluid dark matter (PFDM) distribution. In Schwarzschild-like 
coordinates $(t,r,\theta,\phi)$ the spacetime 
is described by
\begin{equation}
ds^{2}
=
-f(r)\,dt^{2}
+\frac{dr^{2}}{f(r)}
+r^{2}d\theta^{2}+r^{2}\sin^{2}\theta\,d\phi^{2},
\label{metric}
\end{equation}
and the metric function is
\begin{equation}
f(r)
=
1-\frac{2M}{r}
+\frac{\lambda}{r}
\ln\left(\frac{r}{|\lambda|}\right).
\label{metric_function}
\end{equation}
Here, $M$ denotes the black-hole mass parameter, while $\lambda$ 
characterizes the contribution of the surrounding PFDM distribution. 
In geometrized units, $\lambda$ has the dimension of length, so that both 
$\lambda/r$ and the argument of the logarithm are dimensionless. If all 
lengths are normalized by a characteristic scale, for example $M$, the 
corresponding rescaled value of $\lambda$ is dimensionless.

The last term in Eq.~(\ref{metric_function}) represents the modification 
of the Schwarzschild geometry induced by PFDM. In the limit
$\lambda\rightarrow 0$, one has
\begin{equation}
\lim_{\lambda\rightarrow0}
\lambda\ln\left(\frac{r}{|\lambda|}\right)=0,
\end{equation}
and the Schwarzschild metric is continuously recovered,
\begin{equation}
f(r)\rightarrow 1-\frac{2M}{r}.
\end{equation}
Moreover, since
\begin{equation}
\frac{\lambda}{r}
\ln\left(\frac{r}{|\lambda|}\right)
\rightarrow 0,
\qquad r\rightarrow\infty,
\end{equation}
the metric approaches the Minkowski spacetime at large radial distances.
Nevertheless, at finite $r$ the logarithmic contribution modifies the
gravitational field and consequently affects the trajectories and phases
of particles propagating in the vicinity of the compact object.

The position of the event horizon is determined by the positive root of
$f(r_h)=0$, namely
\begin{equation}
r_h-2M+
\lambda\ln\left(\frac{r_h}{|\lambda|}\right)=0.
\label{horizon_eq}
\end{equation}
For the non-negative values of $\lambda$ considered in the present
analysis, the left-hand side of Eq.~(\ref{horizon_eq}) is a monotonically
increasing function of $r$, and therefore the spacetime possesses a
single positive horizon. In the Schwarzschild limit, the usual result
$r_h=2M$ is recovered. Throughout this work, neutrino propagation is
considered outside the event horizon, where $f(r)>0$.

Since our analysis is concerned with weak gravitational lensing, the
relevant trajectories are restricted to the region in which the
gravitational corrections are small. More explicitly, along the
propagation path we require
\begin{equation}
\frac{2M}{r}\ll1,
\qquad
\left|
\frac{\lambda}{r}
\ln\left(\frac{r}{|\lambda|}\right)
\right|\ll1.
\label{weak_field_condition}
\end{equation}
Under these conditions the metric function can be consistently expanded
around its asymptotically flat form. Although the PFDM contribution does
not modify the leading relativistic oscillation phase for purely radial
propagation, as will be shown below, it enters explicitly for non-radial
trajectories through the impact-parameter-dependent part of the phase.
It is therefore the gravitational-lensing configuration that provides
the relevant sensitivity to the parameter $\lambda$ in the neutrino
oscillation probability.

\section{Neutrino oscillations
\label{sec3}}
Neutrinos are produced and detected in flavor eigenstates, denoted by
$|\nu_{\alpha}\rangle$, while their propagation is governed by the corresponding
mass eigenstates $|\nu_i\rangle$. The flavor and mass bases are related through
the leptonic mixing matrix, so that a flavor eigenstate can be written as
\begin{equation}
    |\nu_{\alpha} \rangle=\sum U_{\alpha i}^{*} 
 |\nu_{i} \rangle. 
\end{equation}
Here, $\alpha=e,\mu,\tau$ labels the three flavor eigenstates, whereas
$i=1,2,3$ denotes the mass eigenstates. The matrix $U$ is the
Pontecorvo-Maki-Nakagawa-Sakata (PMNS) leptonic mixing matrix
\cite{Pontecorvo:1957qd, Maki:1962mu, Pontecorvo:1967fh}.

The energy and momentum of each mass eigenstate satisfy the usual mass-shell
relation
\begin{equation}
    E_i^2=\vec{p_i^2}+m_i^2.
\end{equation}
We consider neutrino propagation from a source $S$, located at the spacetime
event $(t_S,x_S)$, to a detector $D$ at $(t_D,x_D)$. The state of the $i$th
mass eigenstate at the detector can then be expressed as
\begin{equation}
    |\nu_i(t_D,x_D) \rangle=\exp(-i \Phi_i)  |\nu_i(t_S,x_S) \rangle, 
\end{equation}
where $\Phi_i$ denotes the phase accumulated during propagation and is given by
$\Phi_i=E_i t-\vec{p_k}* \vec{x}$.
The probability for a neutrino initially produced in the flavor state
$\nu_\alpha$ to be detected as $\nu_\beta$ is therefore
\begin{align}
    &\mathcal{P}_{\alpha \beta}= |\langle \nu_{\beta}|\nu_{\alpha}(t_D,x_D)||^2\nonumber\\&=\sum_{i,j} U_{\beta i}  U_{\beta j}^*  U_{\alpha j}  U_{\alpha i}^* \exp(-i(\Phi_i-\Phi_j)).
\end{align}
In flat spacetime, the phase accumulated by the $i$th mass eigenstate can be
written as
\begin{equation}
    \Phi_i=E_i(t_D-t_S)-p_i(\vec{x}_D-\vec{x}_S).
    \label{eq7}
\end{equation}
A commonly adopted assumption is that the mass eigenstates composing the
initial flavor state are produced with either the same momentum or the same
energy \cite{Bilenky:1978nj}. In addition, we use the ultrarelativistic
approximation
\begin{equation}
    (t_D-t_S)\simeq |\vec{x_D}-\vec{x_S}|,
\end{equation}
which leads to
\begin{equation}
    \Phi_i=(E_i-|\vec{p_i}|) |\vec{x_D}-\vec{x_S}|.
    \label{eq9}
\end{equation}
The energy can be expanded in the relativistic limit as
\begin{equation}
    E_i \simeq E_0+O \left(\frac{m_i^2}{2 E_0}\right),
\end{equation}
and, for $E_i>>m_i$, the phase in Eq.~(\ref{eq9}) takes the form
\begin{equation}
    \Phi_i \simeq \frac{m_i^2}{2 E_0}|\vec{x_D}-\vec{x_S}|,
\end{equation}

\begin{equation}
    \Delta \Phi_{i j} =\Phi_i-\Phi_j \simeq \frac{\Delta m_{i j}^2}{2E_0}(\vec{x}_D-\vec{x}_S),
\end{equation}
where $\Delta m_{i j}^2=m_i^2-m_j^2$, and $E_0$ denotes the average energy of
the relativistic neutrinos emitted by the source. In curved spacetime, the
propagation phase can be written in a covariant form. Accordingly,
Eq.~(\ref{eq7}) is generalized as \cite{Fornengo:1996ef}
\begin{equation}
    \Phi_i=\int_S^D p_\mu^{(i)} d x^\mu,
\end{equation}
where
\begin{equation}
    p_\mu^{(i)}=m_i g_{\mu \nu} \frac{d x^\nu}{ d s},
    \label{momenta}
\end{equation}
is the canonical momentum conjugate to the coordinate $x^\mu$, while $ds$
denotes the spacetime line element and $g_{\mu\nu}$ is the metric tensor.

\section{Neutrino oscillation in black hole immersed by PFDM
\label{sec4}}

Restricting the motion to the equatorial plane, $\theta=\pi/2$, the nonvanishing metric components reduce to
\begin{align}
    &g_{tt}=-f(r),\nonumber\\
    &g_{rr}=\frac{1}{f(r)},\nonumber\\
    &g_{\phi \phi}=g_{\theta \theta}=r^2.
\end{align}
Using these metric components, the canonical momenta defined in Eq.~(\ref{momenta}) can be written as
\begin{align}
    &p_t^{(i)}=-m_i f(r) \frac{dt}{ds}\nonumber,\\
    &p_r^{(i)}=m_i \frac{1}{f(r)} \frac{dr}{ds}\nonumber,\\
    &p_{\phi}^{(i)}=m_i r^2 \frac{d\phi}{ds},
    \label{fmomentum}
\end{align}
where $m_i$ denotes the mass of the $i$th neutrino mass eigenstate and $s$ is the affine parameter along its trajectory. These momentum components satisfy the mass-shell condition
\begin{equation}
    -m_i^2=g^{\mu \nu} p_\mu ^{i} p_\nu ^{i}=-\frac{1}{f(r)} (p_t^{i})^2+f(r) (p_r^{i})^2+\frac{1}{r^2} (p_{\phi}^{i})^2.
\end{equation}
\subsection{Radial propagation}
We first consider radial neutrino propagation, for which $d\phi=0$. 
Using Eq.~(\ref{fmomentum}), the equations of motion can be written as
\begin{equation}
    \frac{d t}{d s}= \frac{p_t}{-m_i f(r)}, 
    \quad 
    \frac{d r}{d s}=\frac{p_r f(r)}{m_i}.
    \label{eqradial}
\end{equation}
Since the spacetime is static, the energy of each neutrino mass eigenstate is conserved. 
We therefore introduce $p_t=-E_k$ and $p_r=p_k(r)$, where the index $k$ labels the corresponding mass eigenstate. 
The phase accumulated by a neutrino propagating radially along the associated null trajectory is then given by
\begin{equation}
    \Phi_k=\int_S^D \left[-E_k \left(\frac{dt}{dr}\right)_0+p_k(r)\right] dr,
\end{equation}
where $S$ and $D$ denote the source and detector, respectively. 
From Eq.~(\ref{eqradial}), the corresponding light-ray relation takes the form
\begin{equation}
    \left(\frac{dt}{dr}\right)_0=\frac{E_0}{p_0(r)} \frac{1}{f(r)^2}.
    \label{eq15}
\end{equation}
Here, $E_0$ denotes the conserved energy of the reference massless particle measured at infinity.

The mass-shell condition gives
\begin{align}
    &p_0(r)=\pm E_0 \frac{1}{f(r)},\\
    &p_k(r)=\pm \sqrt{\frac{1}{f(r)} \left(\frac{1}{f(r)}E_k^2-m_k^2\right)}.
    \label{eq16}
\end{align}
Substituting Eqs.~(\ref{eq15}) and~(\ref{eq16}) into the expression for the phase yields
\begin{equation}
    \Phi_k=\pm \int^D_S \frac{E_k}{f(r)}
    \left[
    -1+\sqrt{1-\frac{m_k^2 f(r)}{E_k^2}}
    \right]dr.
\end{equation}
For ultrarelativistic neutrinos, the quantity $m_k^2 f(r)/E_k^2$ is small, so that the square root can be expanded to leading order. 
The phase then reduces to
\begin{equation}
    \Phi_k=\int^D_S \frac{E_k}{f(r)}
    \frac{m_k^2 f(r)}{2 E_k^2} dr.
\end{equation}
Within the relativistic approximation $m_k\ll E_k$ \cite{Fornengo:1996ef}, one has
\begin{align}
 \label{eq20}
   & E_k \simeq E_0+\mathcal{O}\left(\frac{m_k^2}{2E_0}\right),\\\nonumber
   &E_k \frac{m_k^2}{2 E_k^2}\simeq E_0 \frac{m_k^2}{2 E_0^2}.
\end{align}
The integration between the source and detector therefore gives
\begin{equation}
    \Phi_k\simeq\pm \frac{m_k^2}{2E_0} \int^D_S dr
    =\pm \frac{m_k^2}{2E_0}(r_D-r_S).
\end{equation}
Thus, at the order considered, the explicit dependence on the metric function cancels from the radial propagation phase. 
Consequently, the PFDM correction does not contribute to the leading-order oscillation phase, which coincides with the corresponding Schwarzschild result.
\subsection{Non-radial propagation}
We next consider non-radial neutrino propagation. Using Eqs.~(\ref{fmomentum}) and~(\ref{eq16}), the oscillation phase evaluated along the corresponding light-ray trajectory can be written as
\begin{equation}
    \Phi_k=\int^D_S \left[-E_k\left(\frac{dt}{dr}\right)_0+p_r+J_k\left(\frac{d \phi}{d r}\right)_0\right]dr.
\end{equation}
Since the spacetime is static and spherically symmetric, the energy and angular momentum are conserved along the trajectory. We therefore write
$p_t^{(k)}=-E_k$ and $p_{\phi}^{(k)}=J_k$, where $J_k$ denotes the angular momentum associated with the $k$th neutrino mass eigenstate.
Using the canonical momentum relations along the reference light-ray trajectory, these expressions reduce to
\begin{equation}
    \left(\frac{dt}{dr}\right)_0=\frac{E_0}{p_0(r)} \frac{1}{f(r)^2},\quad 
    \left(\frac{d \phi}{dr}\right)_0=\frac{J_0}{p_0(r)}\frac{1}{r^2 f(r)}.
    \label{eq25}
\end{equation}

The angular momentum $J_k$ may be expressed in terms of the energy $E_k$, the impact parameter $b$, and the asymptotic velocity $v_k^{(\infty)}$ as
\begin{equation}
    J_k=E_k b v_k^{\infty}.
    \label{eq26}
\end{equation}
In the asymptotically flat region, the velocity of the $k$th mass eigenstate is
\begin{align}
    &v_k^{\infty}=\frac{\sqrt{E_k^2-m_k^2}}{E_k} \simeq 1-\frac{m_k^2}{2 E_k^2},\\
    &J_k\simeq E_k b \left(1-\frac{m_k^2}{2 E_k^2}\right),
    \label{eq27}
\end{align}
where the second relation follows from the ultrarelativistic approximation. 
For the corresponding massless trajectory, the angular momentum is simply
\begin{equation}
    J_0=E_0 b.
    \label{eq28}
\end{equation}

Substituting Eqs.~(\ref{eq25})-(\ref{eq28}) into the phase integral gives
\begin{align}
&\Phi_k=\int^D_S\frac{E_0E_k}{p_0(r)f(r)}\\
&\left[-\frac{1}{f(r)}+\frac{p_0(r)p_kf(r)}{E_0E_k }+\frac{b^2}{r^2}\left(1-\frac{m_k^2}{2E_k^2}\right)\right]dr.
\end{align}
The mass-shell condition further yields
\begin{align}
    &\frac{p_0(r)}{E_0 1/f(r)}=\pm \sqrt{1-\frac{b^2}{r^2 \,1/f(r)}},\\
    &\frac{p_0(r) p_k(r)}{E_0 E_k 1/f(r)}=\frac{1}{f(r)}-\frac{b^2}{r^2}-\frac{m_k^2}{2 E_k^2}.
\end{align}
Using these relations, we can rewrite the phase as
\begin{equation}
    \Phi_k=-\int_S^D \frac{E_0}{p_0(r)f(r)}E_k\frac{m_k^2}{2E_k^2}dr.
\end{equation}
Employing the relativistic approximation introduced in Eq.~(\ref{eq20}), we obtain
\begin{align}
    &\Phi_k=-\frac{m_k^2}{2E_0}\int_S^D \frac{E_0}{p_0(r) f(r)}dr=\pm\frac{m_k^2}{2E_0}\int_S^D \frac{1}{\sqrt{1-\frac{b^2f(r)}{r^2}}}dr.
    \label{eq32}
\end{align}

The integral in Eq.~(\ref{eq32}) will be evaluated for two different propagation geometries. 
In the first configuration, the neutrino follows a non-radial trajectory that remains on one side of the gravitational center. 
In the second configuration, the neutrino is emitted from a source $S$, passes the compact object acting as a gravitational lens, reaches a minimum radial distance $r_0$, and subsequently propagates toward the detector $D$. 
The latter configuration is the one relevant for the gravitational-lensing analysis of the oscillation probability.

For the first configuration, and within the weak-field approximation, the quantity entering Eq.~(\ref{eq32}) can be expanded as
\begin{align}
\sqrt{1-\frac{b^2}{r^2}f(r)}
&\simeq
\sqrt{1-\frac{b^2}{r^2}}
\left[
1
+
\frac{Mb^2}{r(r^2-b^2)}
\right.
\nonumber\\
&
\left.
-
\frac{\lambda b^2}{2r(r^2-b^2)}
\ln\!\left(\frac{r}{|\lambda|}\right)
\right].
\end{align}
Substituting this expansion into the phase integral gives
\begin{align}
\Phi_k
\simeq
\pm\frac{m_k^2}{2E_0}
\int_S^D
\Biggl[
&
\frac{1}{\sqrt{1-\frac{b^2}{r^2}}}
-
\frac{Mb^2}
{\sqrt{1-\frac{b^2}{r^2}}\,r(r^2-b^2)}
\nonumber\\
&
+
\frac{\lambda b^2}
{2\sqrt{1-\frac{b^2}{r^2}}\,r(r^2-b^2)}
\ln\!\left(\frac{r}{|\lambda|}\right)
\Biggr]dr.
\end{align}
Performing the integration, one obtains
\begin{multline}
\Phi_k
=
\frac{m_k^2}{2E_0}
\Biggl[
\sqrt{r^2-b^2}
+
\frac{M}{\sqrt{r^2-b^2}}
\\
\shoveleft{
+\frac{\lambda}{2}
\left(
\ln\!\left(r+\sqrt{r^2-b^2}\right)
-
\frac{r}{\sqrt{r^2-b^2}}
\ln\!\left(\frac{r}{|\lambda|}\right)
\right)
\Biggr]_{r_S}^{r_D}
}
\end{multline}
After evaluating the expression at the integration limits, the phase takes the form
\begin{align}
\Phi_k
=
\frac{m_k^2}{2E_0}
\Biggl[
&
\sqrt{r_D^2-b^2}
-
\sqrt{r_S^2-b^2}
\nonumber\\
&
+
M
\left(
\frac{1}{\sqrt{r_D^2-b^2}}
-
\frac{1}{\sqrt{r_S^2-b^2}}
\right)
\nonumber\\
&
+\frac{\lambda}{2}
\Biggl(
\ln\!\frac{r_D+\sqrt{r_D^2-b^2}}
{r_S+\sqrt{r_S^2-b^2}}
\nonumber\\
&
-
\frac{r_D}{\sqrt{r_D^2-b^2}}
\ln\!\left(\frac{r_D}{|\lambda|}\right)
+
\frac{r_S}{\sqrt{r_S^2-b^2}}
\ln\!\left(\frac{r_S}{|\lambda|}\right)
\Biggr)
\Biggr].
\end{align}
In the limit $b\rightarrow0$, this expression reduces to the radial-propagation case discussed above.

We now turn to the lensing configuration. Since the radial momentum changes sign at the point of closest approach, the phase integral is naturally separated into the source-turning-point and turning-point-detector contributions,
\begin{align}
    &\Phi_k(r_S\rightarrow r_0 \rightarrow r_D)=\frac{m_k^2}{2 E_0}\int_{r_0}^{r_S} \frac{1}{\sqrt{1-\frac{b^2 f(r)}{r^2}}}dr\\
    &\nonumber+\frac{m_k^2}{2 E_0}\int_{r_0}^{r_D} \frac{1}{\sqrt{1-\frac{b^2 f(r)}{r^2}}}dr.
\end{align}
The radius of closest approach $r_0$ is determined by the turning-point condition
\begin{equation}
    \left(\frac{dr}{d \phi}\right)_0=\frac{p_0(r_0)r^2 f(r)}{J_0}=0,
\end{equation}
which leads to
\begin{equation}
    1-\frac{b^2}{r_0^2}f(r)=0.
\end{equation}
Since this equation does not admit a simple closed-form solution, we proceed within the weak-field approximation.

For $b\ll r_{S,D}$, expanding in $b/r_S$ and $b/r_D$ and retaining terms up to order $b^2/r_{S,D}^2$, the oscillation phase becomes
\begin{align}
\Phi_k\simeq
\frac{m_k^2}{2E_0}
\Bigg[
&
(r_S+r_D)
\left(
1+\frac{2M}{r_Sr_D}
-\frac{b^2}{2r_Sr_D}
\right)
\nonumber\\
&
+\lambda\ln(2|\lambda|)
\nonumber\\
&
-\frac{\lambda b^2}{8}
\left(
\frac1{r_D^2}
+\frac1{r_S^2}
\right)
\nonumber\\
&
-\frac{\lambda b^2}{4}
\left(
\frac{\ln(r_D/|\lambda|)}{r_D^2}
+
\frac{\ln(r_S/|\lambda|)}{r_S^2}
\right)
\Bigg].\label{eq-phase-nonradial}
\end{align}
This result will be used in the next section to construct the neutrino flavor-transition probability in the gravitational-lensing configuration.

\section{Neutrino oscillation probability
\label{sec5}}

We consider neutrinos propagating in the spacetime described by Eq.~(\ref{metric}). Following previous studies, a neutrino produced at the source $S$ in the flavor state $\nu_\alpha$ and propagating toward the detector $D$ along a trajectory labeled by $p$ can be written as \cite{Swami:2020qdi}
\begin{equation}
    |\nu_{\alpha}(t_D,x_D)\rangle =N \sum_i U_{\alpha i}^* \sum_p \exp(-i \Phi^p_i |\nu_{i}(t_S,x_S)\rangle).
\end{equation}
Here, $\Phi_i^p$ denotes the phase accumulated by the $i$th mass eigenstate along the trajectory $p$, as given in Eq.~(\ref{eq-phase-nonradial}), with $b_p$ representing the corresponding path-dependent impact parameter. The probability for a neutrino produced at the source $S$ in the flavor state $\nu_\alpha$ to be detected at $D$ in the flavor state $\nu_\beta$ is then given by \cite{Chakrabarty:2021bpr}
\begin{align}
\label{eqphase}
    &\mathcal{P}^{lens}_{\alpha \beta}=|\langle\nu_\beta|\nu_{\alpha}(t_D,x_D)||^2\\\nonumber
    &=|N|^2 \sum_{i,j} U_{\beta i} U^*_{\beta j} U_{\alpha j} U^*_{\alpha i} \sum_{p,q} \exp(-i \Delta \Phi^{pq}_{i j}),
\end{align}
where $|N|^2$ denotes the normalization factor,
\begin{equation}
    |N|^2=\bigg(\sum_i |U_{\alpha i}|^2 \sum_{p,q}(-i \Delta \Phi^{pq}_{i j} ) \bigg).
\end{equation}
The phase difference between the $i$th and $j$th mass eigenstates propagating along the trajectories $p$ and $q$, respectively, is defined as
\begin{equation}
   \Delta \Phi^{pq}_{i j}= \Phi^i_p-\Phi^j_q=\Delta m^2_{ij} A_{pq}+\Delta b^2_{pq} B_{ij},
   \label{eq51}
\end{equation}
where
\begin{align}
A_{pq}
&=
\frac{r_S+r_D}{2E_0}
\left[
1+\frac{2M}{r_S+r_D}
-\frac{\Sigma b_{pq}^{2}}{4r_Sr_D}
\right]
+\frac{\lambda}{2E_0}\ln\!\left(2|\lambda|\right)
\nonumber\\
&
-\frac{\lambda\,\Sigma b_{pq}^{2}}{32E_0}
\Bigg[
\frac{1}{r_D^{2}}
+\frac{1}{r_S^{2}}
+2\left(
\frac{\ln\!\left(r_D/|\lambda|\right)}{r_D^{2}}
+
\frac{\ln\!\left(r_S/|\lambda|\right)}{r_S^{2}}
\right)
\Bigg],
\\[3mm]
B_{ij}
&=
-\frac{\Sigma m_{ij}^{2}}{8E_0}
\Bigg[
\frac{1}{r_S}
+\frac{1}{r_D}
+\frac{\lambda}{4}
\left(
\frac{1}{r_D^{2}}
+\frac{1}{r_S^{2}}
\right)
\nonumber\\
&
+\frac{\lambda}{2}
\left(
\frac{\ln\!\left(r_D/|\lambda|\right)}{r_D^{2}}
+
\frac{\ln\!\left(r_S/|\lambda|\right)}{r_S^{2}}
\right)
\Bigg].
\end{align}
The quantities entering these expressions are defined by
\begin{align}
    &\Delta m^2_{ij}=m^2_i-m^2_j,\quad \sum m^2_{ij}=m^2_i+m^2_j,\\\nonumber
    &\Delta b^2_{pq}=b^2_p-b^2_q,\quad
    \sum b^2_{pq}=b^2_p+b^2_q.
\end{align}

\section{Gravitational lensing of neutrinos in spacetime describing black hole solution immersed in perfect fluid dark matter\label{sec6}}

We now simplify the expression for the neutrino flavor-transition probability. Substituting Eq.~(\ref{eq51}) into Eq.~(\ref{eqphase}), we obtain
\begin{align}
    & \mathcal{P}^{lens}_{\alpha \beta}=|N|^2 \bigg[\sum_{i,j} U_{\beta i} U^*_{\beta j} U_{\alpha j} U^*_{\alpha i}\bigg(\sum_{p=q}\exp(-i \Delta m^2_{ij} A_{pp})\\\nonumber
    &+2\sum_{p>q} \cos(\Delta b^2_{pq} B_{ij})\exp(-i \Delta m^2_{ij} A_{pq}) \bigg)  \bigg],
\end{align}
with the normalization factor written as
\begin{equation}
    |N|^2=\bigg(N_{path}+\sum_i |U_{\alpha i}|^2 \sum_{q>p} 2 
 \cos(\Delta b^2_{pq} B_{ii})\bigg)^{-1}.
\end{equation}

For definiteness, we restrict the discussion to neutrino propagation in the equatorial plane, $\theta=\pi/2$, and consider two coherent propagation paths, so that $N_{path}=2$. 
Under these assumptions, the general expression for the transition probability reduces to
\begin{align}
\label{eqprob}
    & \mathcal{P}^{lens}_{\alpha \beta}= |N|^2 \bigg[2 \sum_i |U_{\beta i}|^2 |U_{\alpha i}|^2(1+\cos(\Delta b^2B_{ii}))\\\nonumber
    &+\sum_{i,j \neq i}U_{\beta i} U^*_{\beta j} U_{\alpha j} U^*_{\alpha i}[\exp(-i\Delta m^2_{ij}A_{11})\\\nonumber
    &+\exp(-i \Delta m^2_{ij} A_{22})+2 \cos(\Delta b^2 B_{ij}) \exp(-i \Delta m^2_{ij} A_{12})]\bigg].
\end{align}
The corresponding normalization factor becomes
\begin{equation}
    |N|^2=\bigg(2+2\sum_i |U_{\alpha i}|^2 \cos(\Delta b^2 B_{ii}) \bigg)^{-1},
\end{equation}
where $\Delta b^2=\Delta b^2_{12}$.

In the following, we examine the resulting oscillation probabilities in both two and three flavor frameworks. 
We first consider the simplified two flavor case and then extend the analysis to the full three-flavor scenario. 
The transition probabilities are evaluated within the source-lens-detector plane in the weak-field regime, allowing us to study how gravitational lensing modifies the interference pattern of neutrino mass eigenstates.
\subsection{Two-flavor case}
For the two-flavor transition $\nu_e\rightarrow\nu_\mu$, Eq.~(\ref{eqprob}) reduces to
\begin{align}
     &\mathcal{P}^{lens}_{e \mu}= |N|^2 \sin^2 2\alpha \bigg[\sin^2 \bigg(\Delta m^2 \frac{A_{11}}{2}\bigg)+\sin^2 \bigg(\Delta m^2 \frac{A_{22}}{2}\bigg)\\\nonumber
     &-\cos(\Delta b^2 B_{12}) \cos(\Delta m^2 A_{12})+\frac{1}{2} \cos(\Delta b^2 B_{11})\\\nonumber
     &+\frac{1}{2} \cos(\Delta b^2 B_{22}) \bigg],
\end{align}
where 
\begin{equation}
    |N|^2=[2(1+\cos^2 \alpha \cos(\Delta b^2 B_{11})+\sin^2 \alpha \cos (\Delta b^2 B_{22}))]^{-1},
\end{equation}
and $\Delta m^2=\Delta m^2_{21}$.
\begin{figure*}[ht]
\centering
\begin{minipage}{0.32\textwidth}
  \raggedright
  \label{fig1}
  \small
  Schematic representation of the weak gravitational lensing configuration
considered in this work. Neutrinos emitted by the source $S$ propagate
toward the detector $D$ along trajectories deflected by a static,
spherically symmetric compact object surrounded by perfect fluid dark
matter. The quantities $r_S$ and $r_D$ denote the source-lens and
lens-detector distances, respectively, while $b$ is the impact parameter
and $\phi$ specifies the angular position of the detector.\caption{\label{fig1}}
\end{minipage}%
\hspace{0.02\textwidth}
\begin{minipage}{0.65\textwidth}
  \centering
  \includegraphics[width=\textwidth]{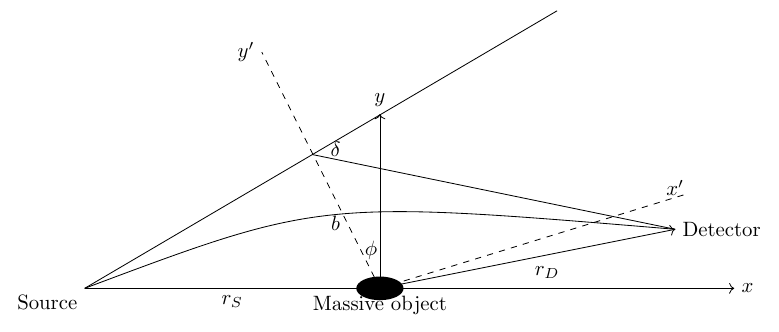}
\end{minipage}
\end{figure*}

Fig.~\ref{fig1} illustrates the weak-lensing geometry adopted in our analysis. 
A neutrino source $S$, a compact object acting as the gravitational lens, and a detector $D$ are assumed to lie in the same plane. 
The lens is described by the static, spherically symmetric black-hole spacetime surrounded by perfect fluid dark matter. 
The source-lens and lens-detector distances are denoted by $r_S$ and $r_D$, respectively, in the $(x,y)$ coordinate system. 
For convenience, we also introduce a rotated coordinate system $(x',y')$, obtained from $(x,y)$ through a rotation by an angle $\phi$,
\begin{equation}
x'=x\cos\phi+y\sin\phi,
\qquad
y'=-x\sin\phi+y\cos\phi.
\end{equation}

In the weak-lensing regime, the neutrino deflection angle $\delta$ is related to the impact parameter $b$ through
\begin{equation}
    \delta \sim \frac{y'_{D} - b}{x'_{D}} = -\frac{4M}{b}=-\frac{2R_{s}}{b},
    \label{eqdelta}
\end{equation}
where $(x'_D,y'_D)$ denotes the detector position in the rotated frame and $R_s=2M$ is the Schwarzschild radius of the lens. 
Using the geometrical relation $\sin\phi=b/r_S$ shown in Fig.~\ref{fig1}, Eq.~(\ref{eqdelta}) can be rewritten as
\begin{equation}
(2R_s x_D + b \, y_D) \sqrt{1 - \frac{b^2}{r_S^2}} = b^2 \Biggl(\frac{ x_D}{r_S} + 1 \Biggr) - \frac{2R_x \, b \, y_D}{r_S}.
\label{13}
\end{equation}
The solutions of Eq.~(\ref{13}) determine the allowed impact parameters for a given source-lens-detector geometry.

For the numerical analysis, we adopt a Sun-Earth-like configuration, with the compact object located at the origin and the detector moving on a circular orbit of radius $r_D$. 
The neutrino source is placed behind the lens and emits ultrarelativistic neutrinos with an average energy $E_0=10~\mathrm{MeV}$. 
The detector coordinates are therefore written as
\begin{equation}
x_D=r_D\cos\phi,
\qquad
y_D=r_D\sin\phi.
\end{equation}
We take
\begin{equation}
r_D=10^8~\mathrm{km},
\qquad
r_S=10^5 r_D.
\end{equation}
For each value of $\phi$, the quartic equation obtained from Eq.~(\ref{13}) is solved numerically, yielding two positive real solutions, $b_1$ and $b_2$, corresponding to the two lensed trajectories connecting the source and detector.

Only solutions satisfying
\begin{equation}
R_x \ll b_p \ll r_D
\end{equation}
are retained in the calculation of the oscillation probability. 
This condition ensures that the neutrino trajectories remain sufficiently far from the lens for the weak-lensing approximation to be valid, while the detector distance remains much larger than the corresponding impact parameters.
 \begin{figure*}[ht]
    \centering
{\includegraphics[width=1\textwidth]{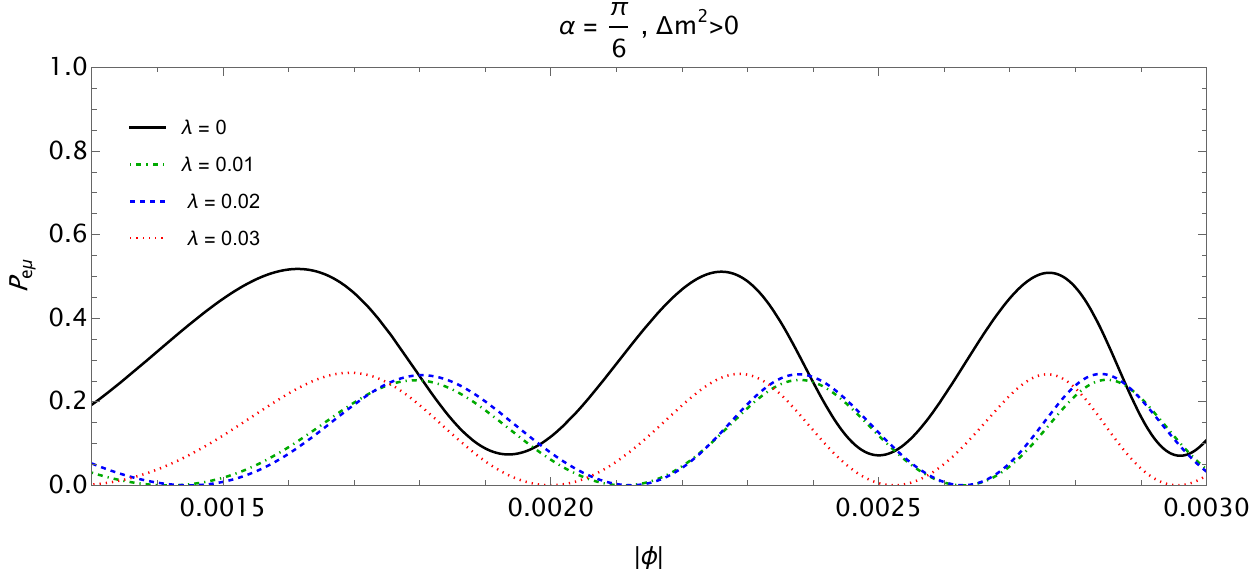}}
{\includegraphics[width=1\textwidth]{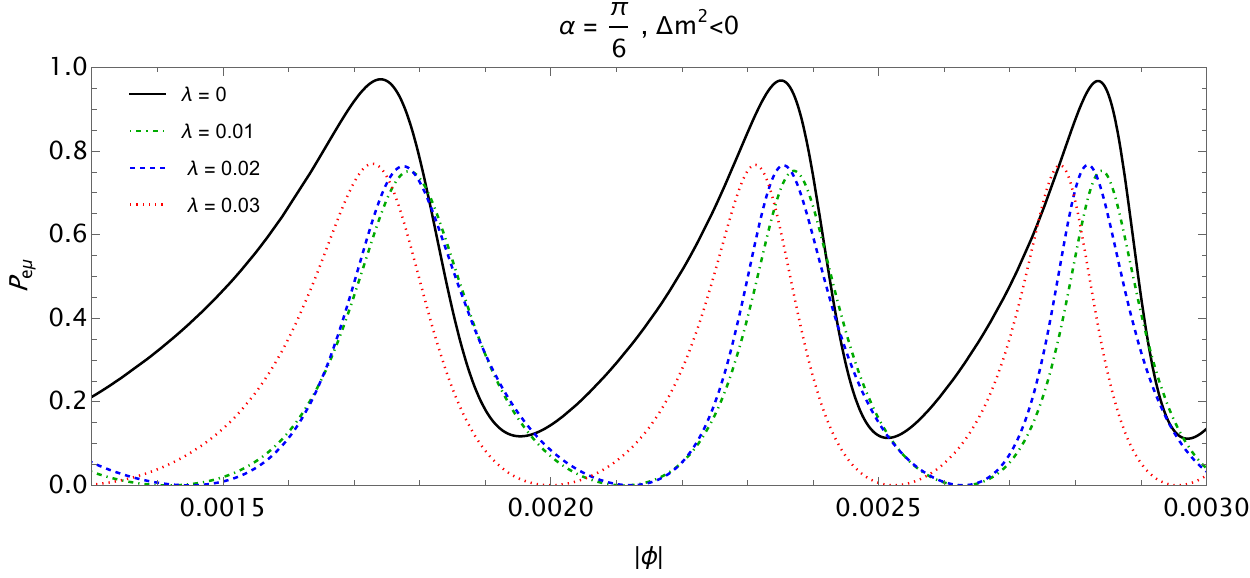}}
\caption{
Two-flavor transition probability $\mathcal{P}_{e\mu}$ as a function of
the absolute angular position of the detector, $|\phi|$, for different
values of the PFDM parameter $\lambda$. The upper panel corresponds to
$\Delta m^2>0$, while the lower panel corresponds to $\Delta m^2<0$.
The curves represent $\lambda=0$ (solid black), $\lambda=0.01$
(dot-dashed green), $\lambda=0.02$ (dashed blue), and $\lambda=0.03$
(dotted red).
}
\label{probabilitypraph}
\end{figure*}
 \begin{figure*}[ht]
    \centering
{\includegraphics[width=\textwidth]{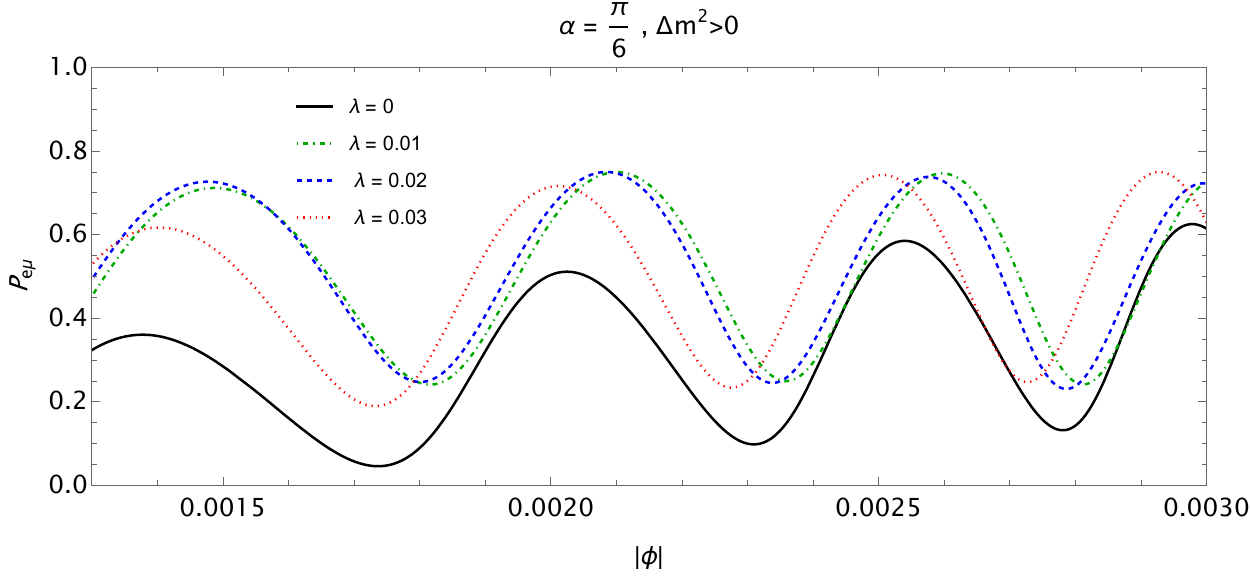}}
{\includegraphics[width=\textwidth]{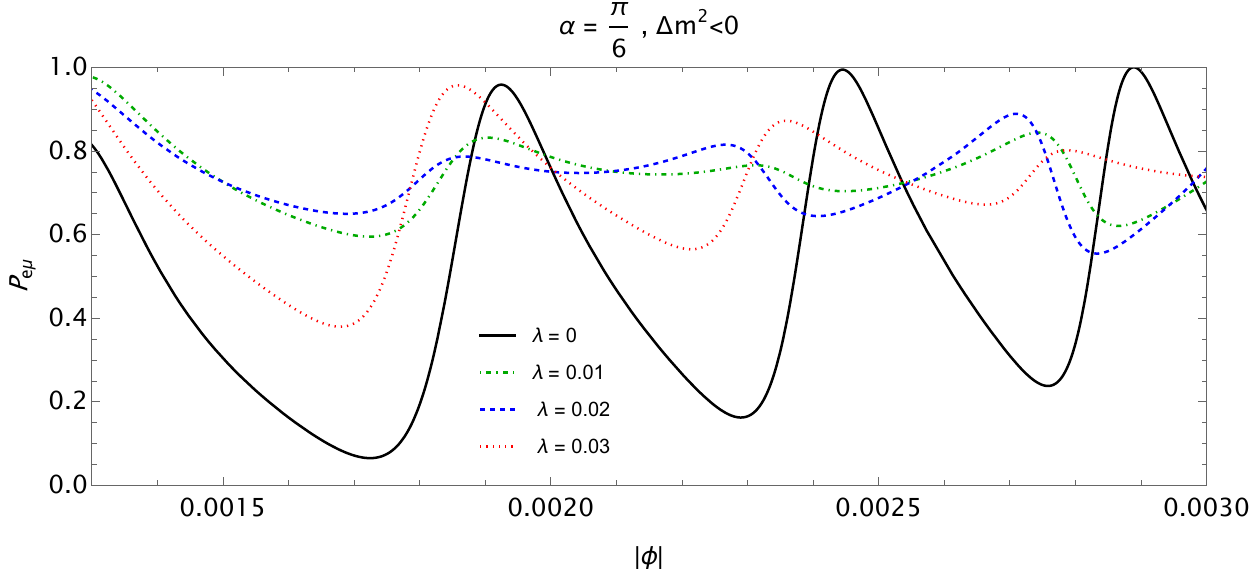}}
\caption{
Two-flavor neutrino transition probability $\mathcal{P}_{e\mu}$ as a
function of $|\phi|$ for different values of the PFDM parameter $\lambda$
and a lens mass $M=7M_{\odot}$. The upper and lower panels correspond to
$\Delta m^2>0$ and $\Delta m^2<0$, respectively. The line styles denote
the same values of $\lambda$ as in Fig.~\ref{probabilitypraph}.
}
\label{7Mgraphs}
\end{figure*}
 
 \begin{figure*}[ht]
    \centering
{\includegraphics[width=1\textwidth]{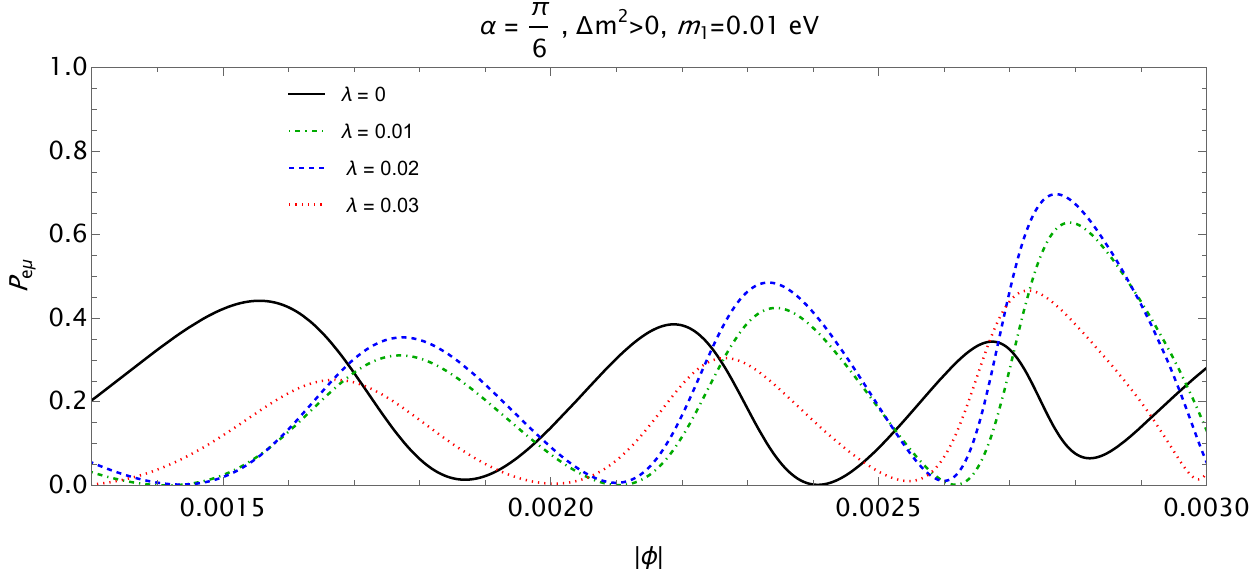}}
{\includegraphics[width=1\textwidth]{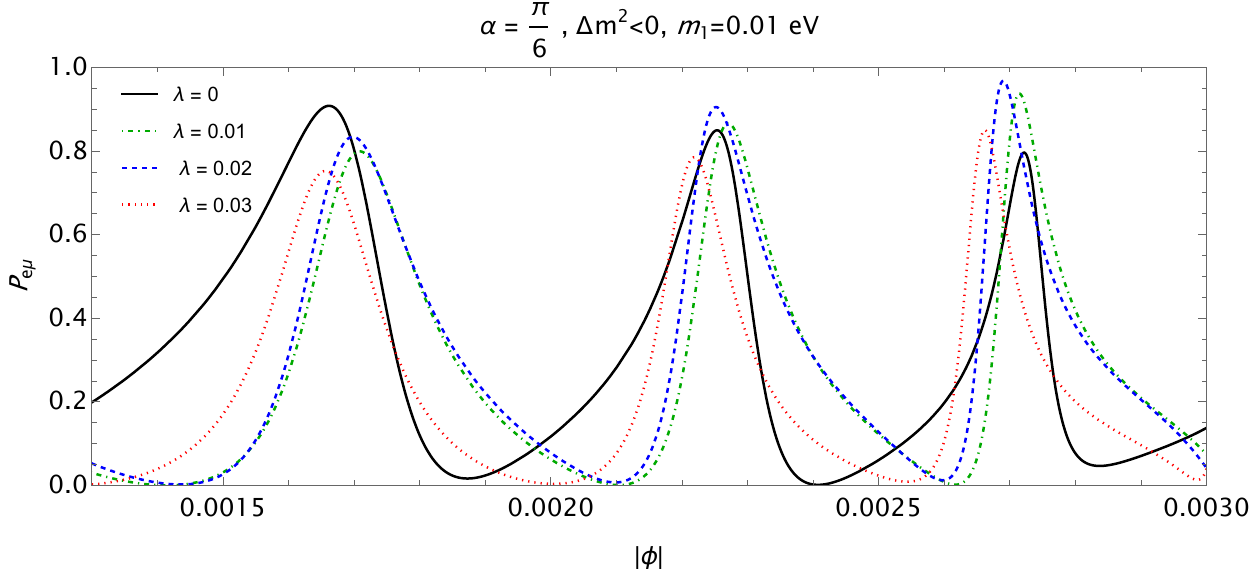}}
\caption{
Two-flavor transition probability $\mathcal{P}_{e\mu}$ as a function of
the detector position $|\phi|$ for a mixing angle $\alpha=\pi/6$ and
lightest neutrino mass $m_1=0.01~\mathrm{eV}$. The upper panel shows the
case $\Delta m^2>0$, whereas the lower panel corresponds to
$\Delta m^2<0$. The curves represent $\lambda=0$ (solid black),
$\lambda=0.01$ (dot-dashed green), $\lambda=0.02$ (dashed blue), and
$\lambda=0.03$ (dotted red).
}
\label{deltam001}
\end{figure*}
 \begin{figure*}[ht]
    \centering
{\includegraphics[width=1\textwidth]{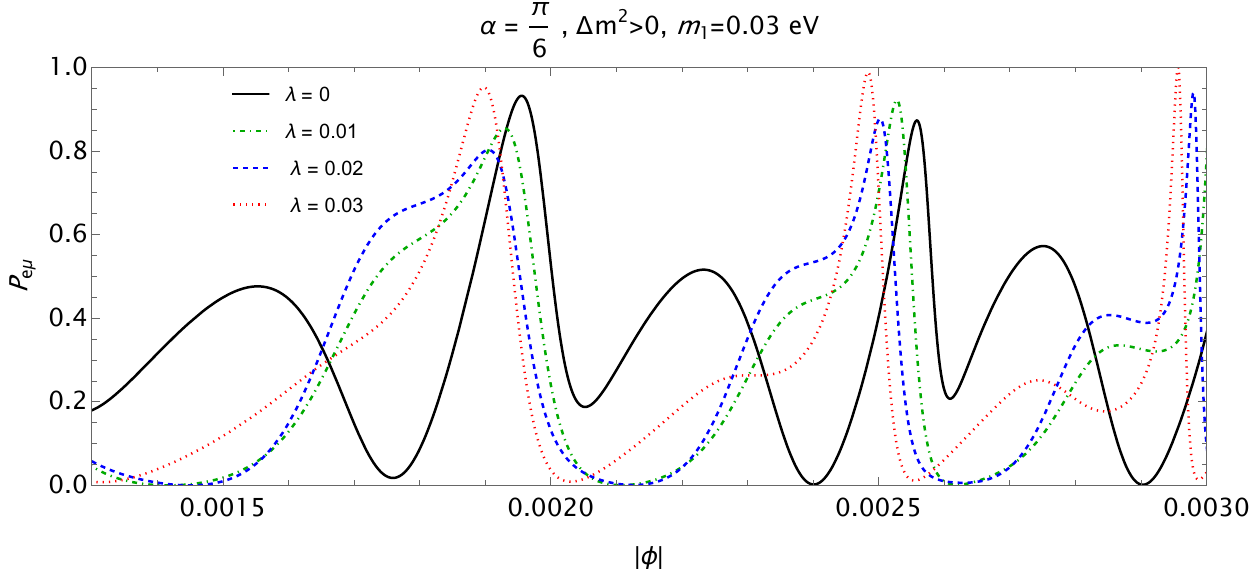}}
{\includegraphics[width=1\textwidth]{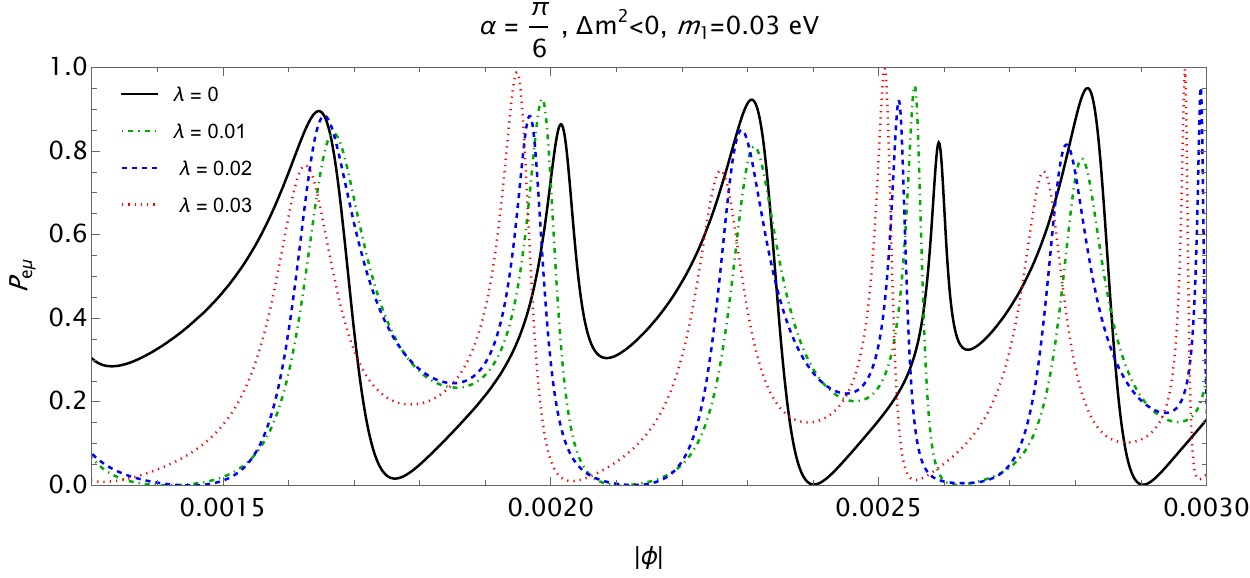}}
\caption{
Two-flavor neutrino transition probability $\mathcal{P}_{e\mu}$ as a function of the absolute detector angle $|\phi|$ for a mixing angle $\alpha=\pi/6$, a lightest neutrino mass $m_1=0.03~\mathrm{eV}$.The upper panel corresponds to $\Delta m^2>0$, while the lower panel represents the case $\Delta m^2<0$.
 The curves represent different values of the PFDM parameter: $\lambda=0$ (solid black), $\lambda=0.01$ (dot-dashed green), $\lambda=0.02$ (dashed blue), and $\lambda=0.03$ (dotted red).
}
\label{deltam003}
\end{figure*}
Fig.~\ref{probabilitypraph} shows the dependence of the two-flavor transition probability $\mathcal{P}_{e\mu}$ on the detector angular position $|\phi|$ for several values of the parameter $\lambda$. For $\Delta m^2>0$, increasing $\lambda$ noticeably modifies both the amplitude and the position of the oscillation maxima. In particular, compared with the $\lambda=0$ case, the nonzero values of $\lambda$ generally suppress the maximum transition probability and introduce a shift of the oscillation pattern along the $|\phi|$ axis. The curves corresponding to $\lambda=0.01$ and $\lambda=0.02$ remain relatively close, whereas $\lambda=0.03$ produces a more pronounced displacement of the maxima.

A similar modification is observed for $\Delta m^2<0$, although the transition probabilities reach considerably larger values in this case. The $\lambda=0$ curve exhibits the highest maxima, while nonzero values of $\lambda$ reduce their amplitudes and shift their locations. These changes indicate that the parameter $\lambda$ affects the phase accumulated by neutrinos along the lensed trajectories and, consequently, changes the interference pattern at the detector.

Fig.~\ref{7Mgraphs} shows the effect of the parameter $\lambda$ on the two-flavor transition probability $\mathcal{P}_{e\mu}$ for $M=7M_{\odot}$. For $\Delta m^2>0$, the $\lambda=0$ curve generally exhibits lower transition probabilities, with pronounced minima between successive oscillation maxima. Introducing a nonzero $\lambda$ increases the overall probability and modifies the phase of the oscillation pattern. In particular, the results for $\lambda=0.01$ and $\lambda=0.02$ remain very close over most of the considered angular interval, whereas $\lambda=0.03$ produces a more visible displacement of the extrema with respect to these two cases. Thus, in the positive-$\Delta m^2$ case, the main effect of $\lambda$ is reflected in both the oscillation amplitude and the position of the maxima and minima along $|\phi|$.

For $\Delta m^2<0$, the influence of $\lambda$ is qualitatively different. The $\lambda=0$ case displays the strongest modulation, characterized by sharp maxima approaching unity and deep minima. For nonzero $\lambda$, these variations become considerably smoother and the probability remains relatively high throughout most of the angular range. The curves for $\lambda=0.01$ and $\lambda=0.02$ show the weakest modulation, while the $\lambda=0.03$ case retains more pronounced oscillatory features. Therefore, the dependence on $\lambda$ is not simply monotonic; rather, changing this parameter modifies both the phase and the contrast of the oscillation pattern, with the response also depending on the sign of $\Delta m^2$.

Figs.~\ref{deltam001} and~\ref{deltam003} illustrate how the neutrino mass parameters affect the angular dependence of the transition probability $\mathcal{P}_{e\mu}$. Compared with the reference case considered previously, the introduction of a nonzero mass scale leads to a progressively stronger modulation of the oscillation pattern. For $m_1=0.01~\mathrm{eV}$, the probability remains relatively smooth for $\Delta m^2>0$, with well-separated maxima and minima over the considered interval of $|\phi|$. The corresponding $\Delta m^2<0$ case exhibits substantially larger conversion probabilities, with several maxima approaching unity. Thus, even for the same value of $m_1$, changing the sign of $\Delta m^2$ produces a pronounced modification of both the amplitude and the phase of the oscillations.

The effect becomes considerably stronger for $m_1=0.03~\mathrm{eV}$. In this case, the oscillation pattern develops sharper and more closely spaced structures, particularly at larger values of $|\phi|$. For $\Delta m^2>0$, several narrow maxima appear and the transition probability for nonzero $\lambda$ can approach unity, in contrast to the smoother behavior obtained for $m_1=0.01~\mathrm{eV}$. For $\Delta m^2<0$, the probability also shows a denser sequence of pronounced maxima and minima, with repeated peaks close to $\mathcal{P}_{e\mu}=1$. This behavior indicates that the oscillation phase is highly sensitive to the mass-dependent contribution entering through $\Delta m^2$.

The comparison further shows that the influence of $\lambda$ becomes more apparent when the mass-dependent phase is larger. While $\lambda$ mainly produces moderate shifts of the extrema for $m_1=0.01~\mathrm{eV}$, the corresponding curves for $m_1=0.03~\mathrm{eV}$ show noticeably larger phase displacements and changes in the shape of individual peaks. Therefore, the effects associated with the spacetime parameter $\lambda$ and the neutrino mass-squared difference are not independent: variations in $\Delta m^2$ change the accumulated oscillation phase and can enhance the sensitivity of the angular probability distribution to the underlying gravitational modification.

\subsection{Three-flavor case}

We now extend the analysis to the complete three-flavor neutrino oscillation framework. The transition probabilities are calculated from Eq.~(\ref{eqprob}) using the $3\times3$ Pontecorvo-Maki-Nakagawa-Sakata (PMNS) mixing matrix. The numerical values of the mixing parameters are taken from the most recent global-fit data. For normal (inverted) ordering, we use $\theta_{12}=33.68^\circ$ ($33.68^\circ$), $\theta_{13}=8.52^\circ$ ($8.58^\circ$), $\theta_{23}=48.5^\circ$ ($48.6^\circ$), $\Delta m_{21}^{2}=7.49\times10^{-5},\mathrm{eV}^{2}$, $\Delta m_{31}^{2}=2.510\times10^{-3},\mathrm{eV}^{2}$ [$\Delta m_{32}^{2}=-2.534\times10^{-3},\mathrm{eV}^{2}$], and $\delta_{\rm CP}=177^\circ$ ($285^\circ$) \cite{Esteban:2024eli}. The source-lens-detector configuration and the remaining geometrical parameters are kept the same as in the two-flavor analysis. In the following, we examine the three independent appearance channels, $P_{e\mu}$, $P_{e\tau}$, and $P_{\mu\tau}$, and determine how their angular profiles change with the spacetime parameter $\lambda$, the lightest neutrino mass, and the mass ordering.

\begin{figure*}[ht]
    \centering
{\includegraphics[width=1\textwidth]{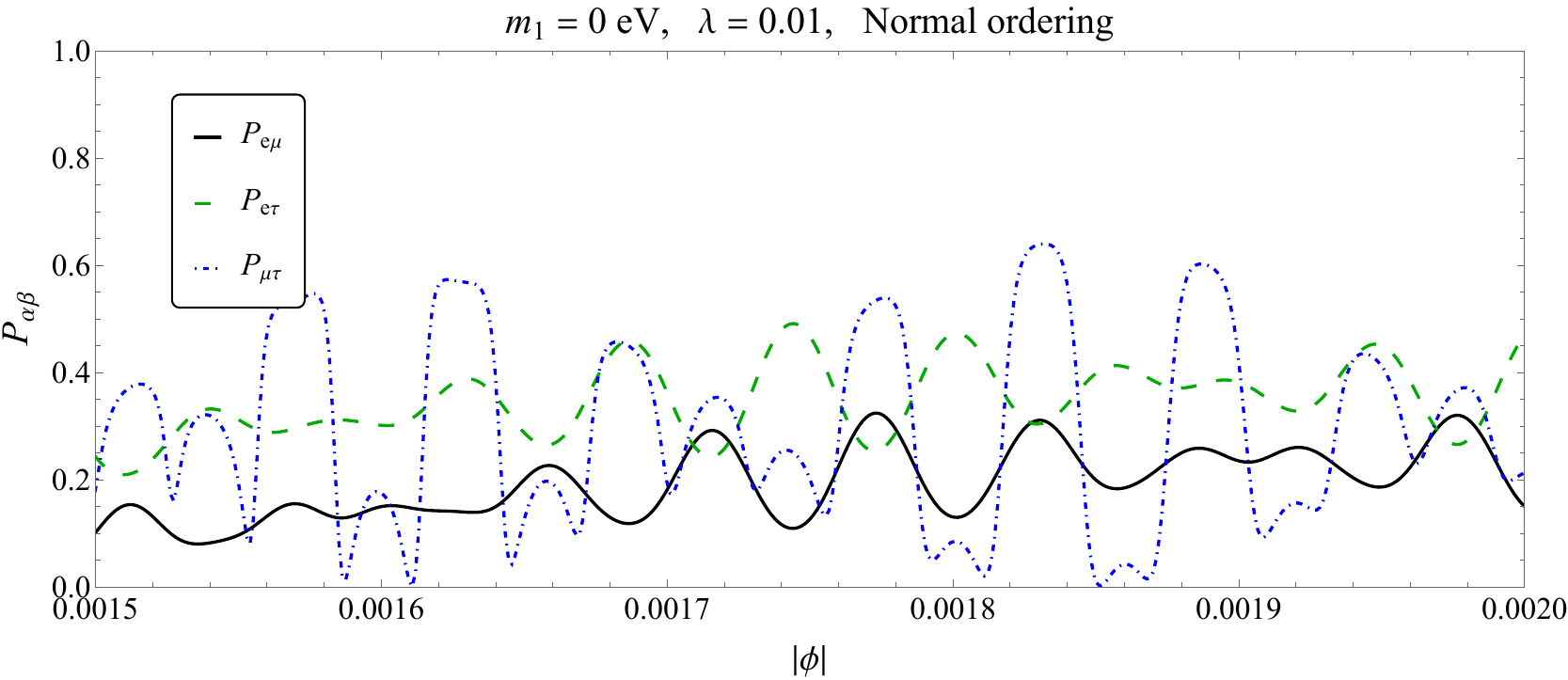}}
{\includegraphics[width=1\textwidth]{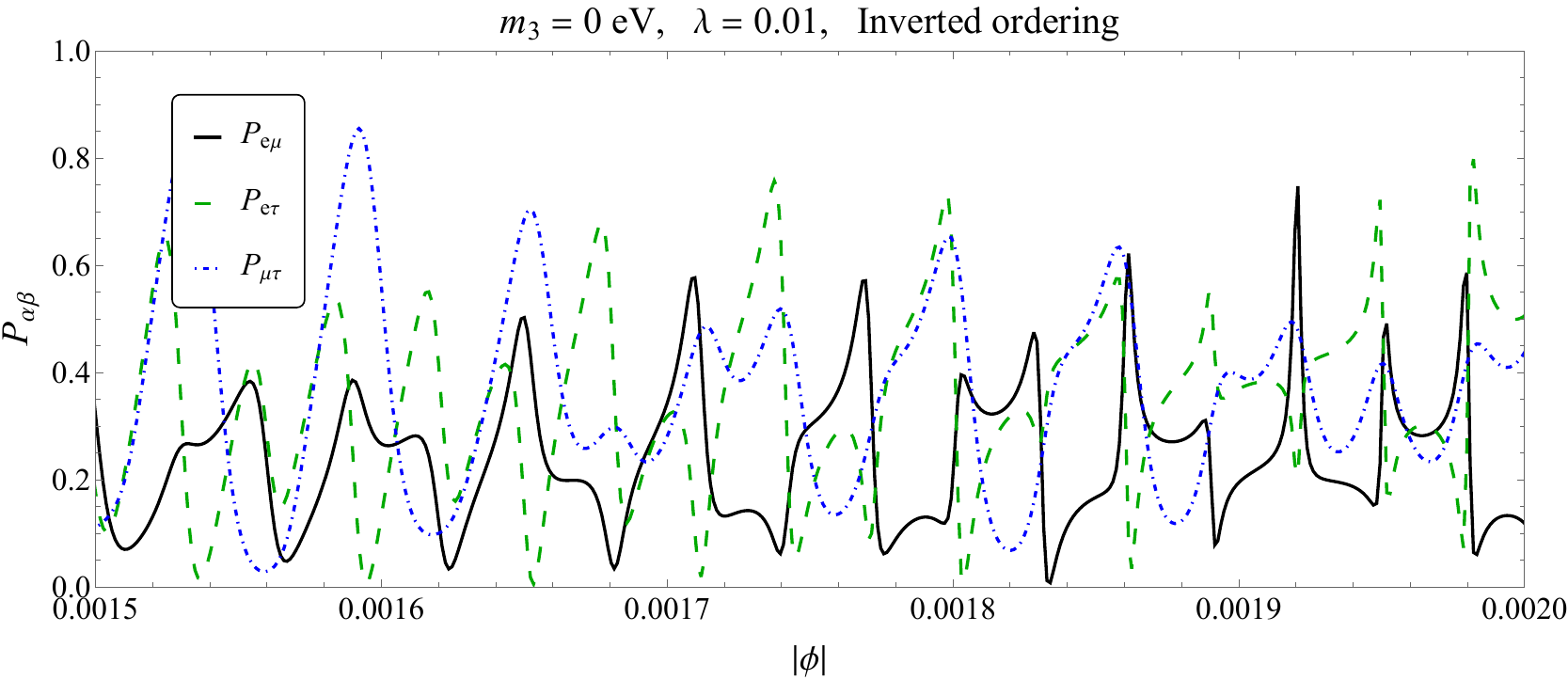}}
\caption{
Three-flavor neutrino appearance probabilities $\mathcal{P}_{\alpha\beta}$ for the channels $\nu_e\rightarrow\nu_\mu$, $\nu_e\rightarrow\nu_\tau$, and $\nu_\mu\rightarrow\nu_\tau$ in the PFDM spacetime with $\lambda=0.01$. The upper row corresponds to normal ordering with $m_1=0$, whereas the lower row shows inverted ordering with $m_3=0$. We use $M=10^3 M_\odot$, $R_s=3~\mathrm{km}$, $r_D=10^8~\mathrm{km}$, $r_S=10^5 r_D$, and $E_0=10~\mathrm{MeV}$. The neutrino mass-squared differences, mixing angles, and Dirac CP phase are taken from the NuFIT 6 global analysis \cite{Esteban:2024eli}.
}
\label{tschw}
\end{figure*}

Fig.~\ref{tschw} provides an overview of the three transition channels. For the normal-ordering configuration with $m_1=0$, the three channels already display clearly different angular structures. In particular, $P_{\mu\tau}$ exhibits the strongest modulation, with repeated maxima extending to considerably larger probabilities than those of the electron-flavor channels. The $P_{e\mu}$ curve remains comparatively low and smooth, whereas $P_{e\tau}$ occupies an intermediate probability range. In the inverted-ordering example, all three channels show a substantially more rapidly varying structure, with narrow maxima and minima distributed throughout the considered interval of $|\phi|$.

\begin{figure*}[ht]
    \centering
{\includegraphics[width=1\textwidth]{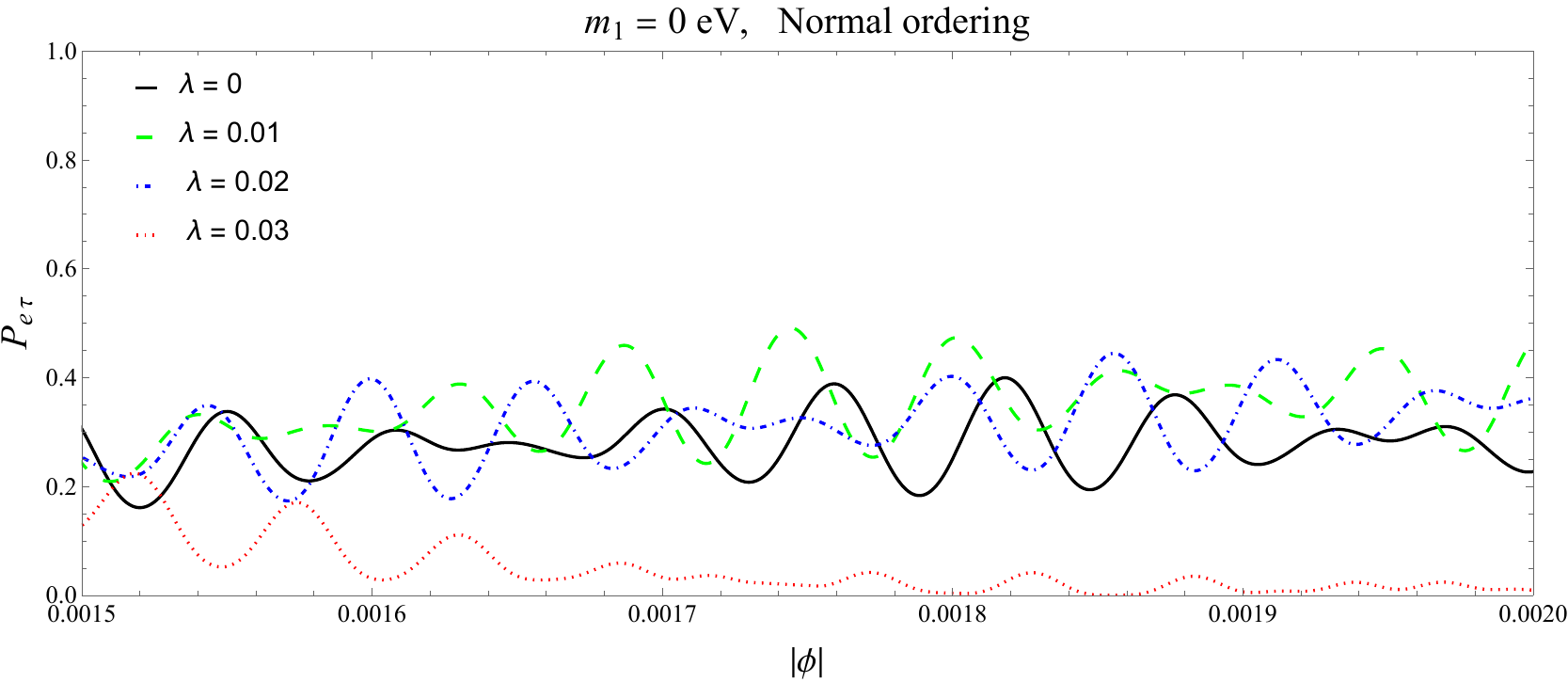}}
{\includegraphics[width=1\textwidth]{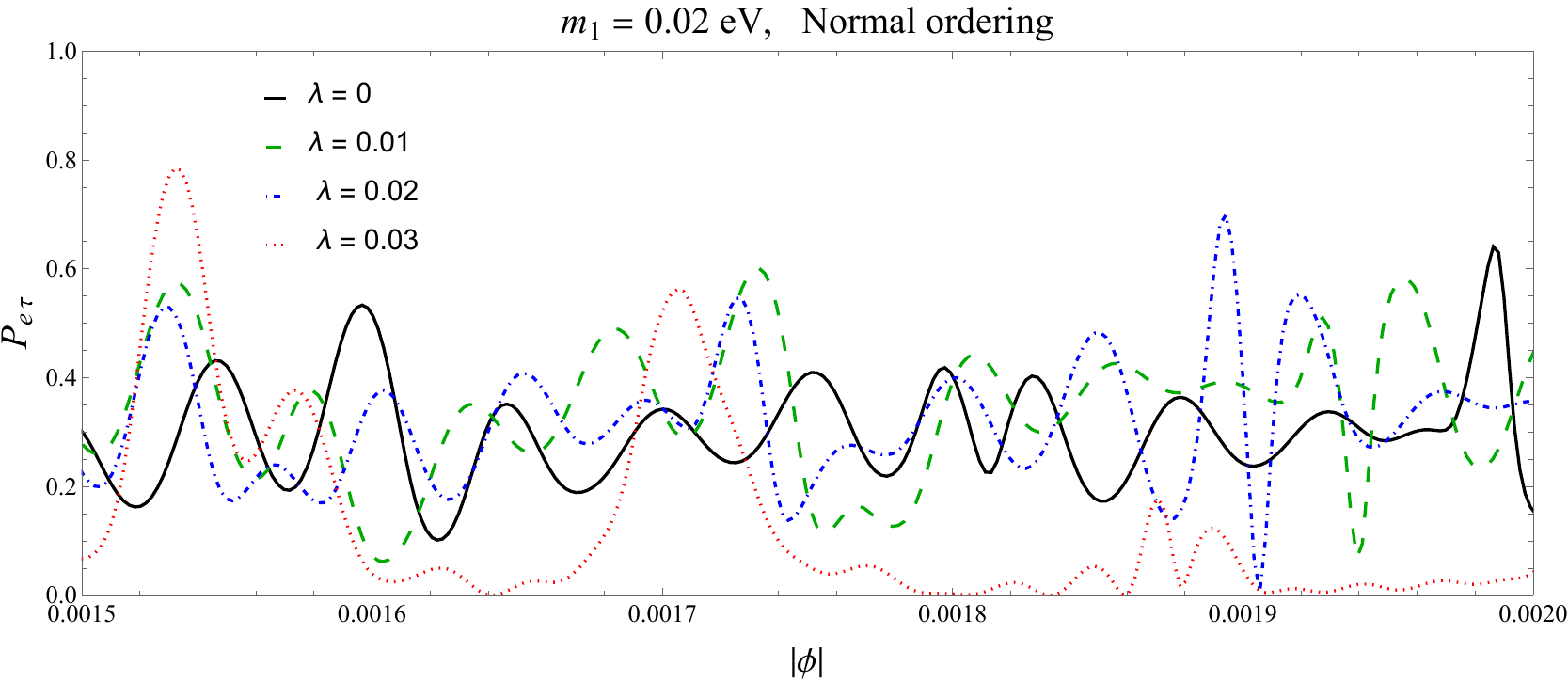}}
{\includegraphics[width=1\textwidth]{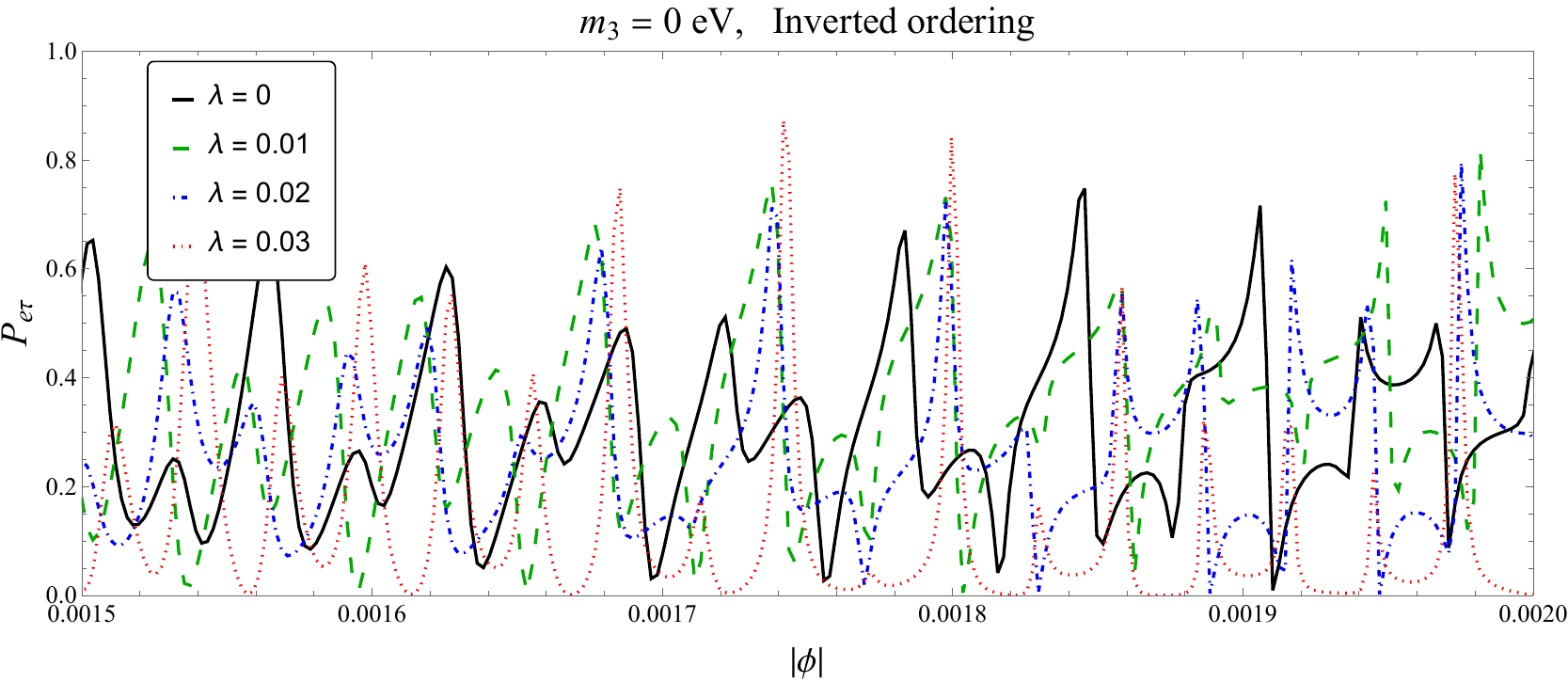}}
\caption{
Three-flavor transition probability $P_{e\tau}$ as a function of
$|\phi|$ for different values of the PFDM parameter $\lambda$. The
upper row corresponds to normal ordering with $m_1=0$, the middle row
to normal ordering with $m_1=0.02~\mathrm{eV}$, and the lower row to
inverted ordering with $m_3=0$. The curves represent $\lambda=0$
(solid black), $\lambda=0.01$ (dashed green), $\lambda=0.02$
(dot-dashed blue), and $\lambda=0.03$ (dotted red).
}
\label{etaugraphs}
\end{figure*}

We next consider separately the dependence of each transition channel on $\lambda$. Fig.~\ref{etaugraphs} shows the probability $P_{e\tau}$. For normal ordering with $m_1=0$, the curves corresponding to $\lambda=0$, $0.01$, and $0.02$ remain within a relatively similar probability range, although changes in the positions and amplitudes of the extrema are clearly visible. A qualitatively stronger modification occurs for $\lambda=0.03$: after the first part of the angular interval, $P_{e\tau}$ is strongly suppressed and remains close to zero over extended regions. The response to $\lambda$ is therefore not monotonic, since the largest value of the spacetime parameter does not simply enhance or uniformly reduce the oscillation amplitude but instead changes the entire angular profile.

The dependence on $\lambda$ becomes considerably more pronounced when the lightest mass is increased to $m_1=0.02,\mathrm{eV}$. In this case, the $P_{e\tau}$ curves develop larger and more irregular variations. The locations of the maxima and minima differ noticeably between the four values of $\lambda$, and several comparatively narrow peaks appear. The $\lambda=0.03$ curve is particularly distinctive, displaying intervals of strong suppression separated by pronounced local maxima. Thus, increasing the lightest mass enhances the separation between the oscillation patterns associated with different values of $\lambda$. For inverted ordering with $m_3=0$, the channel becomes even more strongly structured: all four curves undergo rapid variations over the full angular interval, and changing $\lambda$ produces substantial shifts in both the positions and heights of the individual peaks. No simple ordering of the probabilities with increasing $\lambda$ is observed.

\begin{figure*}[ht]
    \centering
{\includegraphics[width=1\textwidth]{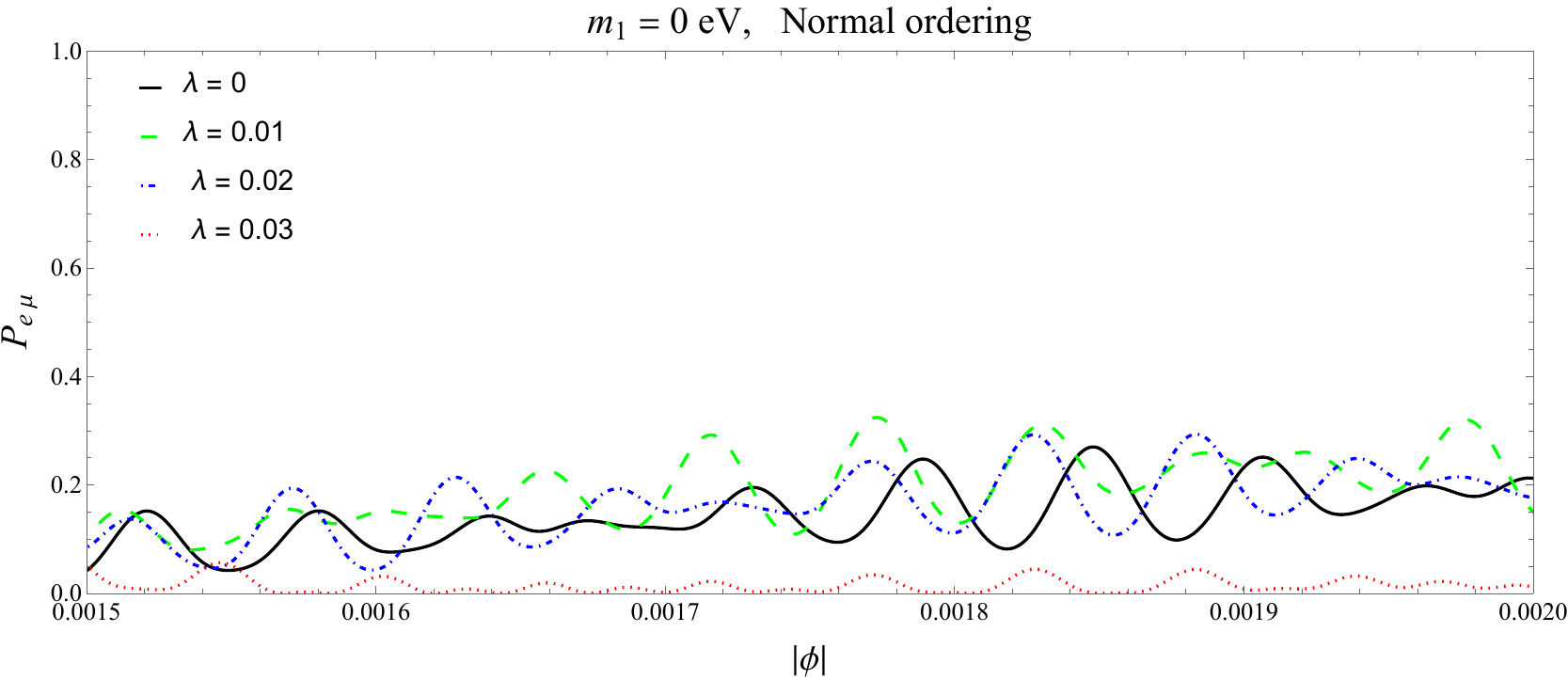}}
{\includegraphics[width=1\textwidth]{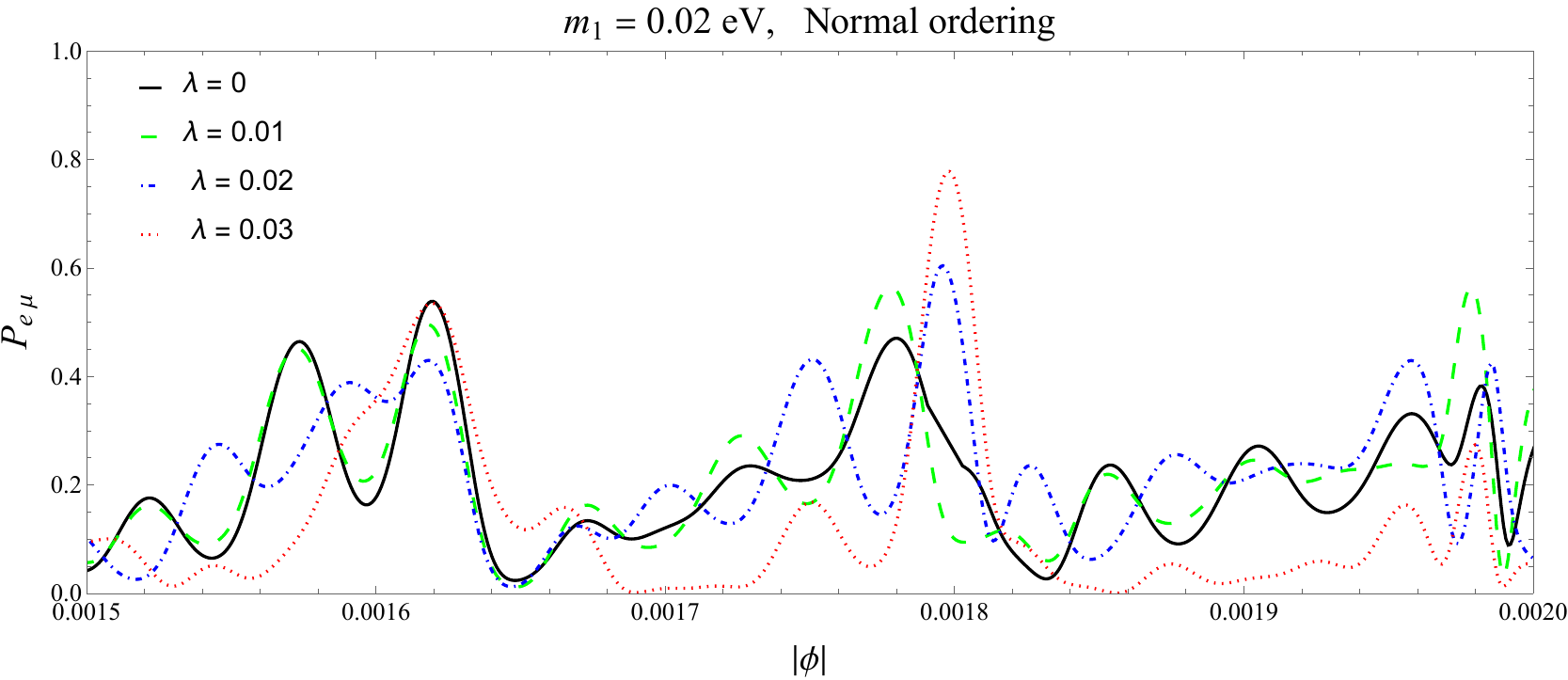}}
{\includegraphics[width=1\textwidth]{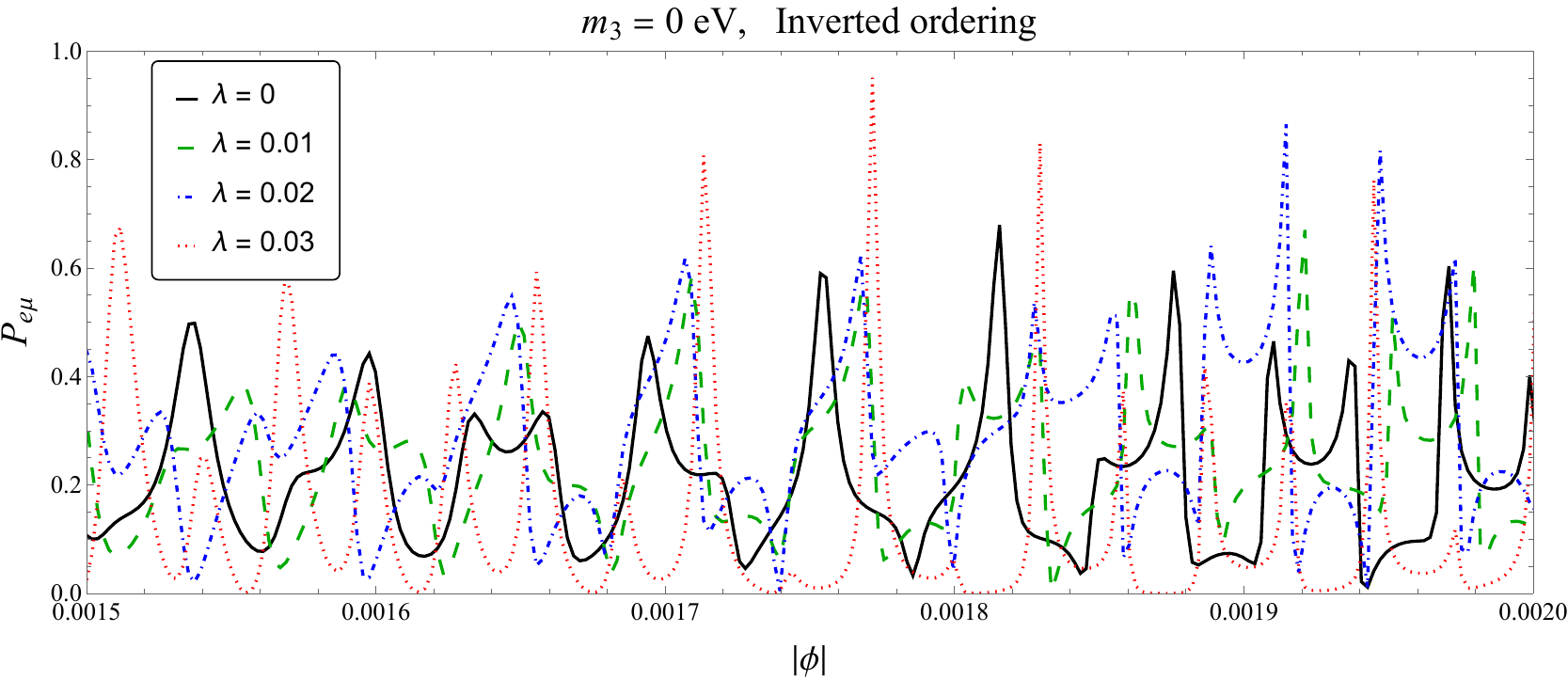}}
\caption{Probability of $\nu_e \rightarrow \nu_\mu$ conversion where the first row shows the case of a normal hierarchy with $m_1=0$\,eV, the middle row shows the case with $m_1=0.02$\,eV, and the last row shows the case with an inverse hierarchy of masses $m_3=0$\,eV. $\lambda=0$ indicates the Schwarzschild case (solid black line), $\lambda=0.01$ (dashed green line), $\lambda=0.02$ (dot-dashed blue line), $\lambda=0.03$ (dotted red line).}
\label{emugraphs}
\end{figure*}

A similar behavior is found for the $P_{e\mu}$ channel, shown in Fig.~\ref{emugraphs}. For normal ordering and $m_1=0$, the transition probability remains comparatively small. The $\lambda=0$, $0.01$, and $0.02$ curves exhibit moderate oscillations, whereas the $\lambda=0.03$ result is strongly suppressed over most of the interval. This suppression is especially evident when compared with the corresponding $\lambda=0.01$ and $\lambda=0.02$ curves, whose maxima reach appreciably larger values.

For $m_1=0.02,\mathrm{eV}$, the $P_{e\mu}$ pattern changes substantially. The curves become more strongly modulated, and their separation with respect to $\lambda$ is considerably enhanced. In particular, the $\lambda=0.03$ case develops a sharp high-amplitude maximum around the central part of the interval, while remaining small over several neighboring angular regions. The curves for the other values of $\lambda$ also exhibit visible phase displacements and changes in peak height. For inverted ordering, $P_{e\mu}$ displays the strongest angular variability of the three cases considered in this figure. The probabilities repeatedly span a large fraction of the allowed interval, and variations of $\lambda$ shift the narrow extrema relative to one another. The effect of $\lambda$ therefore appears mainly through a restructuring of the oscillation pattern rather than through a uniform increase or decrease of the transition probability.

\begin{figure*}[ht]
    \centering
    \includegraphics[width=1\textwidth]{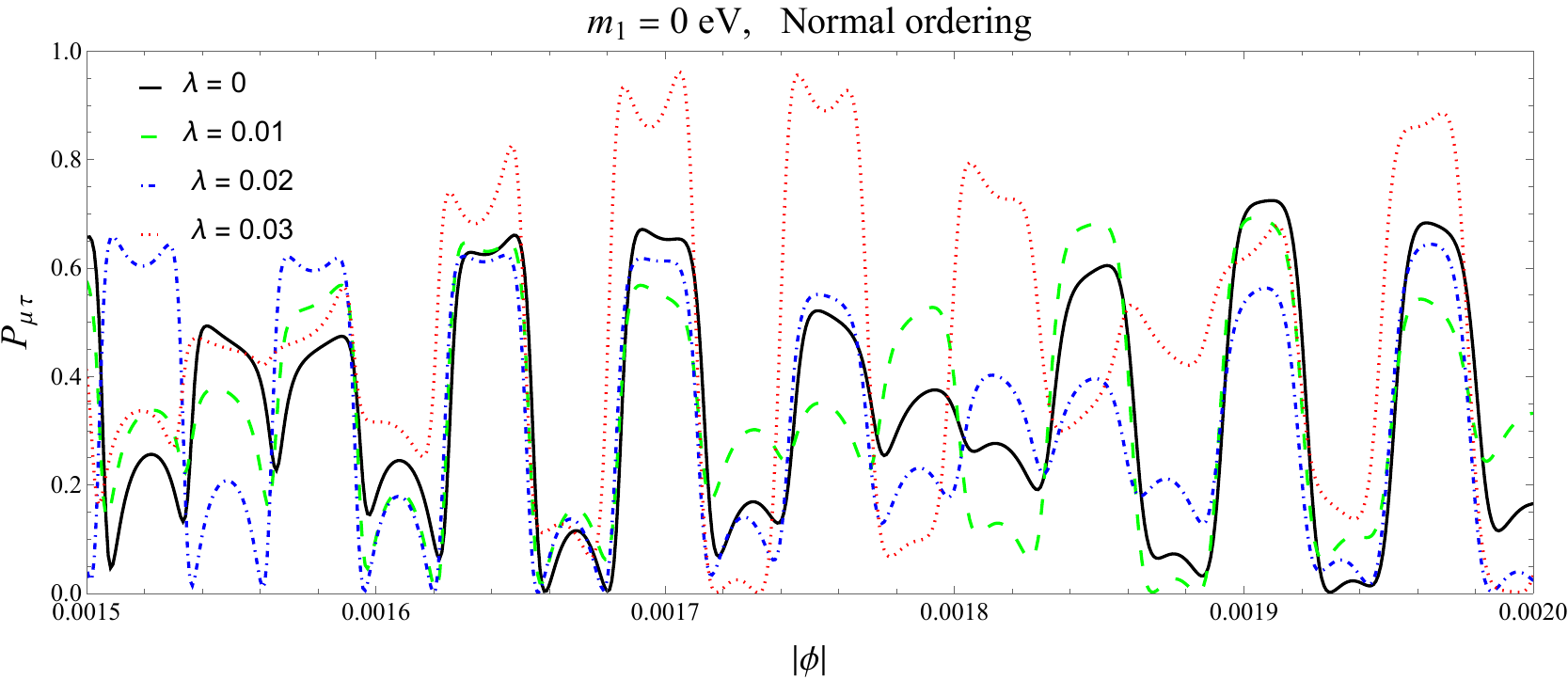}
    \includegraphics[width=1\textwidth]{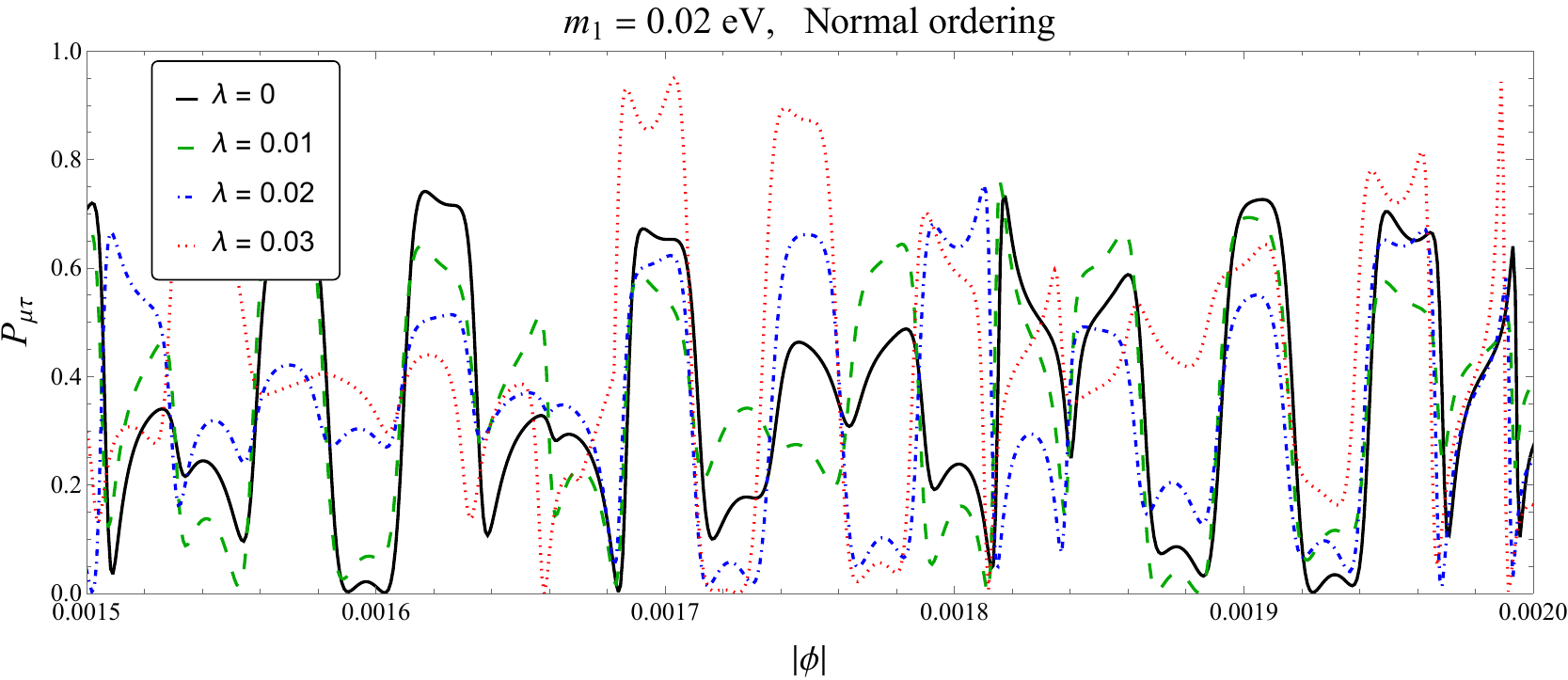}
    \includegraphics[width=1\textwidth]{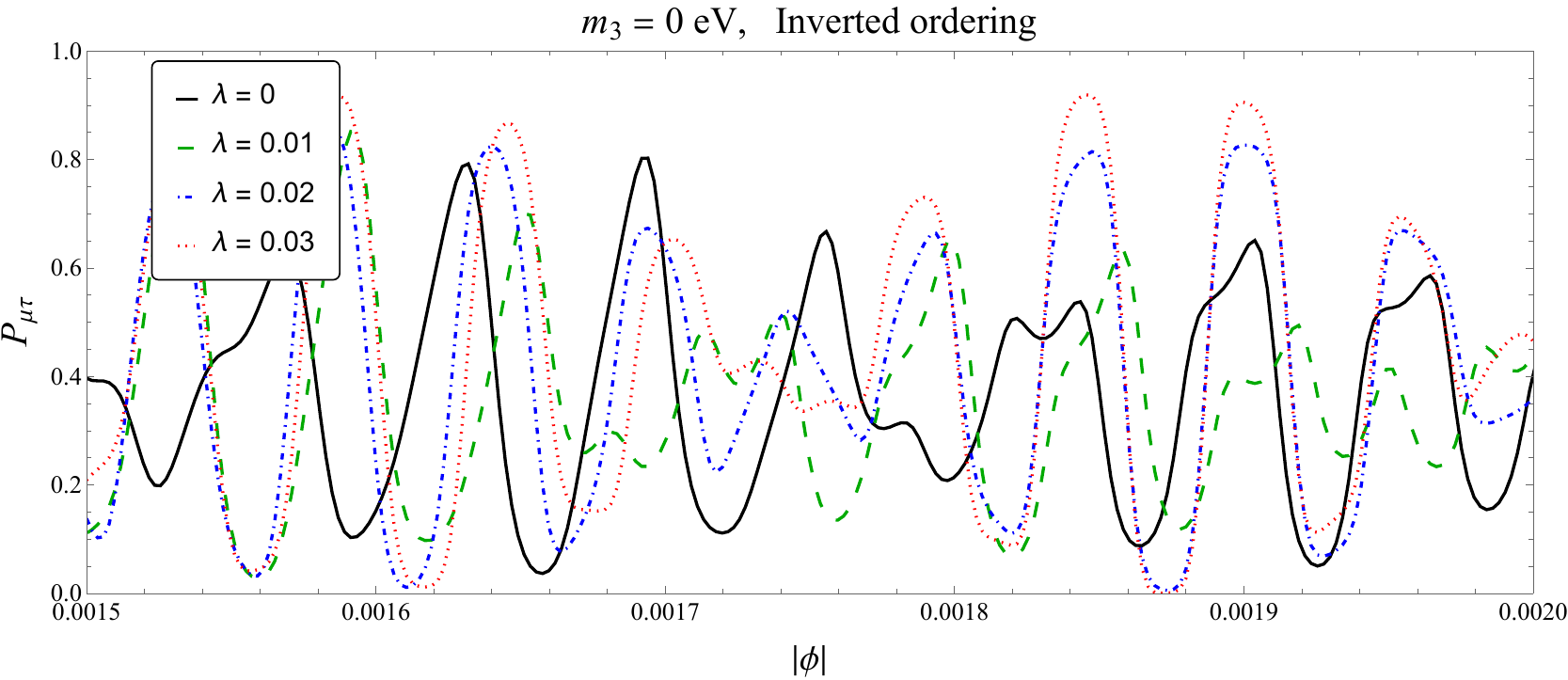}
    \caption{
Transition probability for the $\nu_\mu\rightarrow\nu_\tau$ channel as a function of the detector angular position. The upper row corresponds to normal ordering with $m_1=0$, the middle row to normal ordering with $m_1=0.02~\mathrm{eV}$, and the lower row to inverted ordering with $m_3=0$. The different curves show the dependence on the PFDM parameter, with $\lambda=0$ (solid black), $\lambda=0.01$ (dashed green), $\lambda=0.02$ (dot-dashed blue), and $\lambda=0.03$ (dotted red).
}
    \label{mutau/graph}
\end{figure*}

Fig.~\ref{mutau/graph} presents the corresponding results for $P_{\mu\tau}$. Already for normal ordering with $m_1=0$, $P_{\mu\tau}$ displays large-amplitude oscillations for all considered values of $\lambda$. The $\lambda=0$, $0.01$, and $0.02$ curves follow broadly comparable structures over several angular intervals, although their extrema are displaced. The $\lambda=0.03$ curve shows the most pronounced amplitude modification and reaches values close to unity at several locations. Consequently, the $\mu\rightarrow\tau$ channel is particularly sensitive to variations of the spacetime parameter even when the lightest neutrino mass is zero.

For $m_1=0.02,\mathrm{eV}$, the strong oscillatory character of $P_{\mu\tau}$ is preserved, while the relative positions and amplitudes of the extrema are further modified. Again, the response is not monotonic in $\lambda$: different values of the parameter dominate in different angular regions, and the curves repeatedly cross one another. The $\lambda=0.03$ case nevertheless produces several of the largest peaks and some of the strongest local variations. For inverted ordering with $m_3=0$, the oscillations become more regular than those found for $P_{e\mu}$ and $P_{e\tau}$ in the same ordering. The four curves retain a broadly oscillatory pattern, but changes in $\lambda$ produce clear phase displacements and moderate-to-large modifications of the peak amplitudes.

Taken together, the three-flavor results show that the modification induced by $\lambda$ is strongly channel dependent. For normal ordering with a vanishing lightest mass, $P_{e\mu}$ and $P_{e\tau}$ remain comparatively moderate and exhibit a particularly strong suppression for $\lambda=0.03$, whereas $P_{\mu\tau}$ retains large-amplitude oscillations and becomes enhanced in several angular regions for the same value of $\lambda$. Increasing the lightest mass to $m_1=0.02,\mathrm{eV}$ generally increases the separation between the curves corresponding to different values of $\lambda$, producing stronger phase shifts, sharper local structures, and larger variations of the extrema. In the inverted-ordering case, the dependence on $|\phi|$ is highly structured in all three channels, and variations of $\lambda$ mainly manifest themselves through shifts and reshaping of the oscillation peaks. These results demonstrate that the effect of the spacetime parameter modifies the detailed interference pattern in a manner that depends on the flavor channel, neutrino mass scale, and mass ordering.


\section{Conclusion \label{sec7}}

We have studied neutrino flavor oscillations in the weak-lensing region of a static, spherically symmetric black hole surrounded by perfect fluid dark matter (PFDM). Using the covariant phase integral, we treated radial and nonradial neutrino propagation separately and then constructed the flavor-transition probability for two coherent trajectories connecting the source and detector. This formulation makes it possible to identify which features originate from ordinary flavor oscillations and which arise from interference between the lensed paths.

Accordingly, the PFDM parameter does not enter the radial phase at the order retained in the present expansion. This cancellation is specific to the leading relativistic approximation and should not be interpreted as a general absence of gravitational corrections at higher order. For non-radial propagation, the impact parameter couples the trajectory geometry to the metric function. The weak-field phase consequently contains explicit mass and PFDM terms, including logarithmic contributions proportional to $\lambda$. The lensing configuration is therefore the relevant setting in which the PFDM environment modifies the oscillation phase within our approximation.

For two coherent paths, the phase difference separates into terms proportional to the neutrino mass-squared difference and to the difference between the squared impact parameters. The latter terms involve sums of the individual neutrino mass squares. Thus, when the two paths have unequal impact parameters and remain coherent, the probability is not determined exclusively by $\Delta m_{ij}^{2}$ and may retain sensitivity to the absolute neutrino mass scale. The coefficients multiplying both mass combinations depend on $\lambda$, providing the mechanism by which the PFDM contribution changes the relative phases of the mass eigenstates and the interference between the two trajectories.

In the two-flavor analysis, variations of $\lambda$ shift the angular positions of the maxima and minima and alter their amplitudes and contrast. The effect cannot be represented by a universal enhancement or suppression of the conversion probability: different values of $\lambda$ dominate in different angular intervals, and the curves may repeatedly cross. Increasing the mass-dependent phase produces better angular structures and generally makes the separation among the PFDM profiles more visible. The response also depends on the sign assigned to the mass-squared difference in the two-flavor parametrization, confirming that the principal influence of the environment is exerted through phase interference rather than through an overall probability rescaling.

The three-flavor calculation reveals a further dependence on flavor channel, mass ordering, and the lightest neutrino mass. For the benchmark configurations considered here, the $P_{e\mu}$ and $P_{e\tau}$ channels are comparatively moderate for normal ordering with a vanishing lightest mass, whereas $P_{\mu\tau}$ displays larger modulations and remains sensitive to $\lambda$. A non-zero lightest mass changes the path-interference terms and increases the structural differences among the probability profiles. In the inverted-ordering examples, all three appearance channels exhibit rapidly varying angular patterns, with the PFDM parameter primarily shifting and reshaping the extrema. These numerical results demonstrate a channel-dependent and non-monotonic response to $\lambda$, rather than establishing a simple hierarchy of probabilities with increasing PFDM strength.

The scope of these findings is set by the assumptions of the calculation. We adopted ultrarelativistic neutrinos, a coherent plane-wave treatment, and a weak-field source-lens-detector configuration. Moreover, the impact parameters used in the numerical analysis were obtained from a leading Schwarzschild-type deflection relation. The reported differences therefore isolate the PFDM corrections to the accumulated oscillation phase; a fully self-consistent lensing treatment should also incorporate the $\lambda$-dependent corrections to the deflection angle, image positions, path amplitudes, and magnifications. In an astrophysical application, wave-packet separation, finite source size, a broad neutrino spectrum, and detector energy resolution may reduce or wash out the rapidly varying interference pattern. The present results should consequently be regarded as a theoretical coherent-lensing benchmark rather than a direct observational prediction. Extending the analysis to wave packets, finite coherence lengths, energy and source averaging, PFDM-corrected lens trajectories, and rotating black holes would provide a more realistic assessment of whether neutrino flavor information can constrain the dark-matter environment of compact objects.

\bibliographystyle{apsrev4-2}
\bibliography{cite}
\end{document}